\documentclass[english,amsfonts,amssymb,amsmath,twocolumn,showkeys,nobibnotes,nofootinbib,
groupedaddress,
twoside,aps,pra,reprint]{revtex4-2}
\usepackage{babel}
\usepackage[utf8]{inputenc}
\usepackage{braket}
\usepackage{bbm}
\usepackage{bm}
\usepackage[pdftex, pdftitle={Article}, pdfauthor={Author}, breaklinks=true]{hyperref} 

\usepackage{bookmark}
\hypersetup{colorlinks=true,allcolors=blue}
\usepackage[colorinlistoftodos, color=green!40, prependcaption]{todonotes}
\usepackage{ltablex}
\usepackage{amsthm}
\usepackage{mathtools}
\usepackage{physics}
\usepackage{xcolor}
\usepackage{graphicx}
\usepackage{adjustbox}
\usepackage{placeins}
\usepackage[T1]{fontenc}
\usepackage{lipsum}
\usepackage{csquotes}
\usepackage{svg}
\usepackage{soul}
\usepackage{multirow}
\usepackage{tikz}
\usepackage[normalem]{ulem}
\usepackage{ragged2e} 
\makeatletter
\long\def\@makecaption#1#2{%
  \par
  \vskip\abovecaptionskip
  \begingroup
    \small\rmfamily
    \justifying 
    \noindent #1\@caption@fignum@sep#2\par
  \endgroup
  \vskip\belowcaptionskip
}
\makeatother
\newcommand{\nobracket}{}
\newcommand{\tmmathbf}[1]{\ensuremath{\boldsymbol{#1}}}
\newcommand{\tmop}[1]{\mathrm{#1}}

\renewcommand{\braket}[1]{\left\langle #1 \right\rangle}
\newcommand{\elem}[3]{\left\langle #1 \right| #2 \left| #3 \right\rangle}
\renewcommand{\ket}[1]{\left| #1 \right\rangle}
\newcommand{\Y}{^{89}\mathrm{Y}^+}
\newcommand{\Sr}{^{88}\mathrm{Sr}^+}
\newcommand{\Yb}{^{171}\mathrm{Yb}^+}

\newcommand{\Gauss}{\mathrm{Gauss}}
\newcommand{\nm}{\mathrm{nm}}
\newcommand{\um}{\mathrm{\mu m}}
\newcommand{\mW}{\mathrm{mW}}
\newcommand{\us}{\mu\mathrm{s}}
\newcommand{\ms}{\mathrm{ms}}
\newcommand{\s}{\mathrm{s}}
\newcommand{\Hz}{\mathrm{Hz}}
\newcommand{\kHz}{\mathrm{kHz}}
\newcommand{\MHz}{\mathrm{MHz}}
\newcommand{\GHz}{\mathrm{GHz}}
\newcommand{\THz}{\mathrm{THz}}
\newcommand{\Tesla}{\mathrm{T}}
\newcommand{\meter}{\mathrm{m}}

\newcommand{\ssOneSZero}{5s^2\,{}^1S_0}
\newcommand{\dsThreeDOne}{4d5s\,{}^3D_1}
\newcommand{\dsThreeDTwo}{4d5s\,{}^3D_2}
\newcommand{\dsThreeDThree}{4d5s\,{}^3D_3}
\newcommand{\dsOneDTwo}{4d5s\,{}^1D_2}
\newcommand{\ddThreeFTwo}{4d^2\,{}^3F_2}
\newcommand{\ddThreeFThree}{4d^2\,{}^3F_3}
\newcommand{\ddThreeFFour}{4d^2\,{}^3F_4}

\newcommand{\ddThreePOne}{4d^2\,{}^3P_1}
\newcommand{\ddThreePTwo}{4d^2\,{}^3P_2}
\newcommand{\ddOneDTwo}{4d^2\,{}^1D_2}
\newcommand{\ddOneGFour}{4d^2\,{}^1G_4}
\newcommand{\spThreePZero}{5s5p\,{}^3P_0}
\newcommand{\spThreePOne}{5s5p\,{}^3P_1}

\newcommand{\dpThreeFFour}{4d5p\,{}^3F_4}
\newcommand{\dpThreeDThree}{4d5p\,{}^3D_3}

\begin{document}

\title{Yttrium ion as a platform for quantum information processing}
\author{Christopher N. Gilbreth}
\email{christopher.gilbreth@quantinuum.com}
\affiliation{Quantinuum, 303 South Technology Ct., Broomfield, CO 80021, USA}

\author{Dmytro Filin}
\author{Marianna S. Safronova}
\affiliation{Department of Physics and Astronomy, University of Delaware, Newark, DE, USA}  

\author{Guanming Lao}
\affiliation{Department of Physics and Astronomy, University of California, Los Angeles}
\author{Eric R. Hudson}
\affiliation{Department of Physics and Astronomy, University of California, Los Angeles}
\affiliation{Center for Quantum Science and Engineering, University of California, Los Angeles, California 90095, USA}
\affiliation{Challenge Institute for Quantum Computation, University of California, Los Angeles, California 90095, USA}

\begin{abstract}
Engineering large-scale quantum computers which simultaneously provide high-fidelity quantum operations, low memory errors, low crosstalk, and reasonable resource usage remains an outstanding challenge across quantum computing platforms. In trapped ions, progress has largely focused on alkaline-earth and ytterbium ions, whose simple electronic structures facilitate control over their internal state. Here we investigate singly-ionized yttrium ($\Y$), a two-valence-electron ion whose ground-state manifold hosts a nuclear-spin qubit and which also features a variety of low-lying metastable manifolds, for applications in quantum information processing.  Because experimental data are limited, we perform high-resolution laser-induced fluorescence spectroscopy to measure the hyperfine structure of several low-lying levels, and carry out comprehensive electronic structure calculations to determine lifetimes, transition matrix elements, and hyperfine coefficients for manifolds addressable with visible, near-visible, or infrared wavelengths. Using these results, we analyze schemes for qubit storage, initialization, readout, leakage mitigation, and single- and two-qubit gates. These results position $\Y$ as a uniquely capable next-generation trapped-ion qubit, combining field-insensitive nuclear-spin or clock-qubit storage with spectrally isolated transitions for operations.
\end{abstract}

\maketitle
\section{Introduction}
Trapped ions are a leading platform for quantum information processing, allowing for coherence times many orders of magnitude longer than the time required for individual quantum operations, as well as the highest fidelity quantum gates~\cite{Gaebler2016, ballance2016high, Clark2021, hughes2025}, state preparation and measurement (SPAM)~\cite{Christensen20, An2022}, and integrated system performance~\cite{Moses2023, Ransford2025}. 
This performance can be attributed in part to the relative simplicity and reproducibility of atomic structure, which allows for accurate calculations and high-fidelity control.
To date, most trapped ion qubits have been realized in ionized alkaline earth atoms, whose single valence electron gives rise to a relatively simple, `alkali-like' level structure characterized by low-lying $^2S$ and $^2P$ manifolds with an additional low-lying $^2D$ manifold resulting from nearby $d$ orbitals.
Ytterbium ions, also commonly used, exhibit a similar level structure with an additional low-lying $^2F$ manifold resulting from 
excitation of an electron from the otherwise closed $f$ shell.

Despite this success, it is interesting to consider the question: \textit{can the performance of a trapped ion quantum computer be improved by choosing a more complex atomic ion to host the qubit?}
Several considerations suggest that the answer may indeed be affirmative.
First, as a trapped ion processor is scaled it is important to avoid qubit crosstalk, where electromagnetic fields intended for one ion have unintended effects on another.
As proposed in the \emph{omg} protocol, a more complex atomic structure provides the opportunity for effectively hiding quantum information from fields that may cause crosstalk~\cite{Allcock2021}. 
Second, despite the field-leading coherence times of trapped ions, 
small coherent $Z$ errors can accumulate to an appreciable level during long circuits due to, e.g., uncontrolled magnetic fields~\cite{Moses2023}.
A more complex atomic structure can be expected to host a variety of field-insensitive (so-called clock) states for robust quantum information storage. 
Third, more complex structures allow for use of advanced operational protocols, such as \textit{omg}~\cite{Allcock2021}, hosting multiple qubits per atom~\cite{Campbell2022poly}, and quantum error correction via monolithic codes~\cite{Jain2024, DeBry2025Error,Aydin2025, Sumin2025}. 

A natural choice for exploration of more complex atomic structure is to move by one column in the periodic table to Group III (scandium group).
Ions of these atoms have the potential to have alkaline-earth-like structure, which despite their increased complexity are known to allow relatively simple laser cooling and quantum operations.
Upon ionization, Sc assumes a non-alkaline-earth-like [Ar]$3d4s$ electronic configuration leading to $^3D_1$ ground state manifold, while Y$^+$ assumes an alkaline-earth-like [Kr]$5s^2$ electronic configuration leading to $^1S_0$ ground state configuration. 
Further, $\Y$ has nuclear spin $I = 1/2$, meaning that \textit{a qubit can be hosted in the ground state in a purely nuclear degree of freedom}, a fundamentally new feature compared to group II ions. Since the nuclear spin exhibits roughly 2000$\times$ lower intrinsic sensitivity to magnetic perturbation than a spin-1/2 electronic degree of freedom, this provides a nearly-ideal qubit storage manifold. 

Qubits hosted in a nuclear spin degree of freedom have been demonstrated in neutral atoms, including ${}^{171}\mathrm{Yb}$ and ${}^{87}\mathrm{Sr}$, with long coherence times and a full suite of quantum operations achieved~\cite{Jenkins2022, Ma2022, Barnes2022}. 
${}^{171}\mathrm{Yb}$ is unique among stable group II atoms for its spin-1/2 nucleus, providing ideal two-state nuclear-spin qubit manifolds, with zero electronic angular momentum, in both ground ($6s^2\,{}^1S_0$) and metastable ($6s6p\,{}^1P_0$) levels. 
$\Y$ is similarly unique among stable group III ions for its spin-1/2 nucleus, providing a similarly ideal qubit in its ground state manifold ($\ssOneSZero$). 
A host of metastable states also arise from the low-lying $4d5s$ and $4d^2$ configurations in $\Y$, providing a rich landscape for designing quantum information processing operations.

The electronic structure of $\Y$ has not been widely studied. 
Basic information on some (but not all) important dipole transitions are currently available in public databases~\cite{NIST_ASD}, and a few hyperfine coefficients for metastable manifolds have been measured~\cite{wnnstrm_high_1994}, while higher-order transition rates, the majority of hyperfine coefficients, and metastable lifetimes are generally unknown. 

Therefore, here we perform high-resolution spectroscopy measurements and \emph{ab initio} electronic structure calculations to characterize properties of $\Y$ that are relevant to quantum information processing.
We measure the hyperfine structure of relevant low-lying and excited-state manifolds, and in particular measure the hyperfine coefficient of the excited $\spThreePOne$ level\footnote{Note in this work we will follow NIST designations for electronic levels, but will often omit spectroscopic prefix, e.g., "z" in $5s5p\,z\,^3P_1$ in the text for brevity. Additionally we follow NIST conventions for terms "level" and "state" as in Ref.~\cite{NIST_handbook}.}. 
Our electronic structure calculations include transition matrix elements, hyperfine coefficients, and lifetimes for the large number of levels below or near the energies of the lowest-lying dipole-accessible excited manifolds ($5s5p$ and $4d5p$), which are addressable with visible, near-visible, or infrared wavelengths.
Based on these results, we propose and analyze several schemes for quantum operations, including qubit storage, state initialization and readout, leakage mitigation, and single- and two-qubit gates.

Our results show that $\Y$ offers an expanded operational toolkit for trapped‑ion quantum computing. Its nuclear spin ground-state qubit provides a new type of field-insensitive storage manifold for a trapped ion, while multiple long-lived metastable manifolds host finite-field clock qubits and nearly-closed cycling transitions suitable for measurement operations that are spectrally isolated from stored qubits. Its level structure allows for a wide variety of gate implementations, including natural options for a variety of laser-based and magnetic gradient-based two-qubit gates. This combination of field-insensitive storage, spectrally isolated operations, and flexible gate modalities positions $\Y$ as a uniquely promising platform for large-scale, low-error trapped-ion quantum processors.

The remainder of the paper is outlined as follows. 
In Sec.~\ref{sec:level_structure}, we discuss the level structure and key features of $\Y$, as well as report experimental measurements of the hyperfine structure of the $\dsThreeDOne$, $\dsOneDTwo$, and $\spThreePOne$ levels using high-resolution laser-induced fluorescence spectroscopy. 
In Sec.~\ref{sec:electronic_structure}, we present electronic structure calculations predicting lifetimes, hyperfine coefficients, and transition matrix elements for the low-lying energy levels. 
A complete table of transitions matrix elements for levels up through $\dpThreeDThree$ is presented in Appendix~\ref{appendix:electronic_structure}. 
In Sec.~\ref{sec:quantum_ops} we discuss the implementation of quantum operations on $\Y$ for use in quantum information processing, before concluding in Sec.~\ref{sec:conclusion}. 
Additional details of the theory used in this analysis are given in Appendix~\ref{appendix:quantum_ops}. 

\section{Level structure and key features}\label{sec:level_structure}

\begin{figure*}[t]
    \centering
    \begin{tikzpicture}
        \node[anchor=south west,inner sep=0] (img) at (0,0)
            {\includegraphics[width=\textwidth]{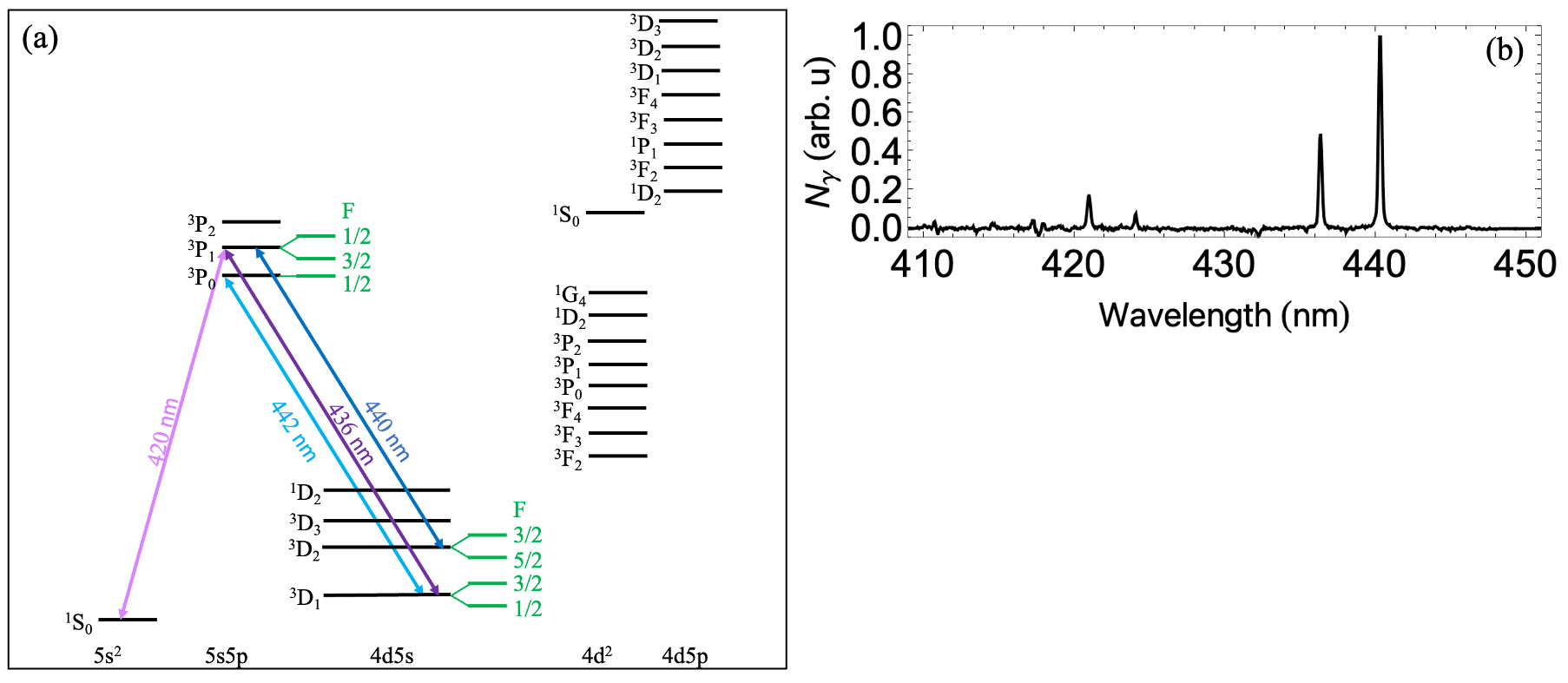}};
        \begin{scope}[x={(img.south east)},y={(img.north west)}]
        \node[anchor=west] at (0.96,0.49) {(c)};
            \node[anchor=west] at (0.52,0.22) {
                \begin{tabular}{ccccc}
                    \hline\hline
     \begin{tabular}{c}
         \text{Final}\\ \text{states}
     \end{tabular} &\begin{tabular}{c}
         \text{Transition}\\ \text{wavelengths}
     \end{tabular}& \begin{tabular}{c}
         \text{Measured}\\ \text{Relative} \\ \text{Strength}
     \end{tabular}& \begin{tabular}{c}
         \text{Theory (\%)}\\ \text{}
     \end{tabular}& \begin{tabular}{c}
         \text{Ref.(\%)}\\ \text{\cite{NIST_ASD, hannaford1982oscillator}}
     \end{tabular} \\
    \hline
       $5s^2~^1S_0$ & 420.6nm &   0.18(13)  &  10.8(20)  & $11.3(1)$ \\
       $4d5s~^3D_1$ & 435.9nm &  0.48(13) &  26.72(24) & $28.6(3)$ \\
       $4d5s~^3D_2$ & 439.9nm & 1 &  55.53(86) & $59.3(3)$ \\
       $4d5s~^1D_2$ & 488.3nm &  -       &  0.67(6)   & $0.77(6)$ \\
       $4d^2~^3P_0$ & 1010.8nm&  -       &  2.07(17)  & - \\
       $4d^2~^3P_2$ & 1033.3nm&  -       &  2.69(25)  & - \\
    \hline\hline
                \end{tabular}
            };
        \end{scope}
    \end{tikzpicture}
    \caption{\label{fig:level_diagram_combo} (a) 
    Low-lying levels of $\Y$ considered in this work (not to scale). (b) DLIF spectra of the $4d5s~^3D_2 - 5s5p~^3P_1^o$ 440~nm transition, the three largest peaks from left to right correspond to decay to the $5s^2~^1S_0$, $4d5s~^3D_1$ and $4d5s~^3D_2$ states, respectively. 
    The spectral contamination between these lines can be attributed to fluorescence of neutral $\mathrm{Y}$ excited by the ablation process. 
    The spectrometer calibration typically has $\approx 1$~nm accuracy.
    (c) Branching ratios of the $5s5p~^3P_1^o$ decays measured in this work and comparison to both previous results~\cite{NIST_ASD, hannaford1982oscillator} and the theory reported here.
    Theoretical results are obtained from the rates presented in Tab.~\ref{trans_tab}.
    }
\end{figure*}

Fig.~\ref{fig:level_diagram_combo}a shows the low-lying electronic energy levels of $^{89} \text{Y}^+$.  
Here, the electron configurations $5 s^2$, $4 d 5 s$, $4 d^2$, $5 s 5 p$, and a selection of levels with configuration $4 d 5 p$, are included.
The ion exhibits some similarity to singly-ionized lutetium ($\mathrm{Lu}^+$)~\cite{Kaewuam2020}, but with the $4 d^2$ levels occurring at lower energy than $5 s 5 p$, introducing a large number of spectator levels below the first dipole transition ($\lambda \approx 421~\nm$) from the ground state manifold, $\ssOneSZero \leftrightarrow \spThreePOne$.

The low-lying $\dsThreeDOne$ level is expected to be long-lived due to selection rules, which forbid $E 1$ and $E 2$ transitions to the ground states, allowing only for a highly suppressed $M 1$ transition. 
For comparison, the corresponding $5 d 6 s \;^3 D_1$ level in $^{175} \mathrm{Lu}^+$ has a calculated lifetime of $1.9 \times 10^5 \text{s}$~\cite{Paez2016}. 
Below, we predict a lifetime of $4.5 (5) \times 10^{10} \text{s}$ for this level. We calculate shorter but still relatively long lifetimes for the higher-energy $4d5s$ levels, ranging from $\sim 5\times 10^3\,\s$ to $\sim 100\,\s$. Calculated lifetimes for all manifolds up through $\dpThreeDThree$ are given in Sec.~\ref{sec:electronic_structure}.

The nuclear spin $I = 1 / 2$ provides a potentially ideal nuclear-spin storage qubit defined in the ground-state manifold $\ssOneSZero$. 
With zero electronic angular momentum and a relatively small nuclear magnetic moment $\mu_I = -0.137298(5)\mu_N$~\cite{Stone2019} ($\mu_N$ is the nuclear magneton), the magnetic field sensitivity of $209.5\,\Hz/\Gauss$ is significantly less than, e.g., the $M = 0$ ground-state qubit of $\Yb$ when operated at $4\,\Gauss$ ($2.5\,\kHz/\Gauss$). 
Further, AC Zeeman shifts from oscillating magnetic fields parallel to the quantization axis have no appreciable effect, while shifts from fields fields perpendicular to the quantization axis are only significant if the oscillation frequency is near the nuclear Zeeman splitting. 

Due to the weak coupling of the nuclear spin to external fields, it is difficult to perform quantum logic operations directly on the nuclear spin. 
One option to facilitate such operations is to coherently shelve the qubit into a metastable manifold.  This is discussed in more detail in Sec.~\ref{sec:shelving}.
The lack of spectator states in the ground-state manifold allows leaked population in metastable levels to be deterministically flushed into the qubit subspace, providing a relatively simple way to mitigate leakage during computation.

The lowest-energy metastable manifold, $\dsThreeDOne$, has only six states and a nearly closed cycling transition to the $\spThreePZero$ manifold with wavelength $\lambda \approx 442~\nm$, which is forbidden at leading order from decaying to the $\ssOneSZero$ ground states. In Sec.~\ref{sec:electronic_structure}, we analyze the hyperfine quenching of the $\spThreePZero$ manifold to determine that its branching fraction outside of this cycling transition is $\sim 2\times10^{-9}$, sufficiently small to enable high-fidelity readout.
Using this transition for readout, while storing spectator qubits in the ground-state manifold, would effectively eliminate crosstalk during measurement.

Additionally, we identify a potential cycling transition between $\dpThreeFFour$ and $\dsThreeDThree$ which could be used for measurement that is spectrally isolated from both the $\ssOneSZero$ and the $\dsThreeDOne$ manifolds. This is discussed in Sec.~\ref{sec:measurement}.

\subsection{Laser spectroscopy of cryogenically cooled $\Y$}
To better characterize this electronic structure we produced cryogenically cooled $\Y$ by laser ablation of a $\mathrm{Y}$ target into a cryogenic buffer gas cell filled with Ne at 20~K and recorded the spectroscopy of select transitions via both total and dispersed laser-induced fluorescence~\cite{Zhu2022} using a narrow-band, frequency-doubled, continuous-wave Ti:Sapphire laser. 
The ablation process was found to provide significant population in both the $5s^2$ and $4d5s$ manifolds, allowing direct interrogation of transitions originating from these levels. 
Therefore, we first recorded dispersed laser-induced fluorescence (DLIF) after excitation on the transitions at roughly 421~nm, 436~nm, 440~nm and 442~nm, as depicted in Fig~\ref{fig:level_diagram_combo}(a).
The recorded DLIF spectra following excitation on these transitions are identical and their sum is shown in Fig.~\ref{fig:level_diagram_combo}(b), with the three most significant features being decays, from left to right, to the $5s^2~^1S_0$, $4d5s~^3D_1$ and $4d5s~^3D_2$ states.
Relative to the strongest decay to the $4d5s~^3D_2$, the measured decay strengths to the $5s^2~^1S_0$ and $4d5s~^3D_1$ states were found to be $0.18(13)$ and $0.48(13)$, respectively.
These ratios were found to be consistent with both previous measurements~\cite{NIST_ASD, hannaford1982oscillator} and the theoretical calculations performed here.

\begin{figure}[h!]
    \centering
    \includegraphics[width = 0.9\columnwidth]{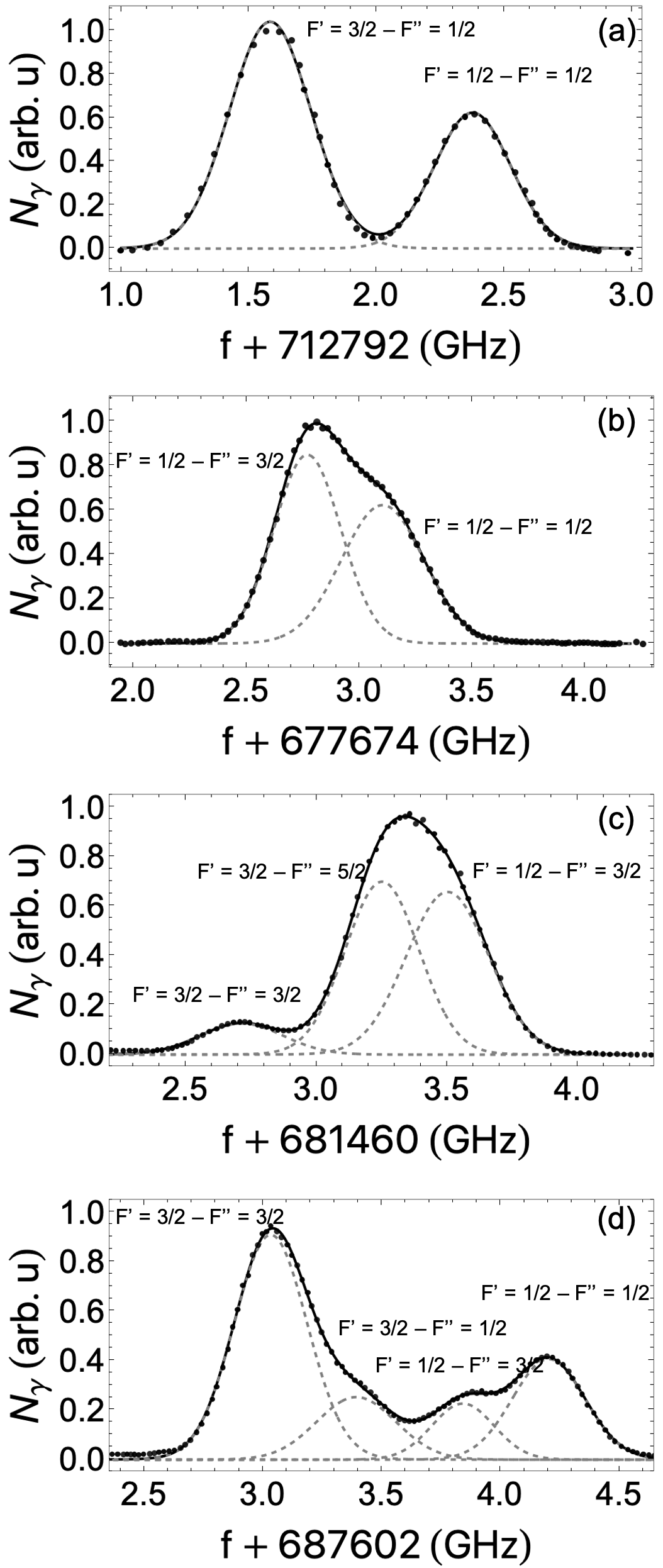}
    \caption{LIF spectra for (a) $5s5p~^3P_1^o \leftrightarrow 5s^2~^1S_0$, (b) $5s5p~^3P_0^o \leftrightarrow 4d5s~^3D_1$, (c) $5s5p~^3P_1^o \leftrightarrow 4d5s~^3D_2$, and (d) $5s5p~^3P_1^o \leftrightarrow 4d5s~^3D_1$ transitions shown in Fig.~\ref{fig:level_diagram_combo}. 
    The recorded number of detected photons is plotted versus laser excitation frequency.
    Each spectrum is fit with a multi-component Voigt profile, shown individually as a dashed curve, while the total profile is shown as a solid line.
     }
    \label{fig3:Yttrium_data}
\end{figure}

To characterize the hyperfine structure of these levels, high resolution spectroscopy using the total laser-induced fluorescence (LIF) was recorded and shown in Fig.~\ref{fig3:Yttrium_data} for the (a) $5s5p~^3P_1^o \leftrightarrow 5s^2~^1S_0$, (b) $5s5p~^3P_0^o \leftrightarrow 4d5s~^3D_1$, (c) $5s5p~^3P_1^o \leftrightarrow 4d5s~^3D_2$, and (d) $5s5p~^3P_1^o \leftrightarrow 4d5s~^3D_1$ transitions.
The width of the observed features corresponds to Doppler broadening for a temperature of $\sim 50$~K.
While higher than the than the 20~K buffer gas temperature, this result is reasonable given buffer gas cooling times and the fact that ablation produces ions with temperatures $> 1000$~K. 

The hyperfine constants, presented in Tab.~\ref{tab:hfconst}, were determined from these data by fitting the spectra with a sum of Voigt probability distributions representing the individual hyperfine resolved transitions -- labeled by the upper and lower state total angular momentum $F'$ and $F''$, respectively.
Several relevant features appear in this data. 
The two hyperfine lines $F'=\frac{3}{2}, \frac{1}{2} \leftrightarrow F''=\frac{1}{2}$ of the $5s5p~^3P_1^o \leftrightarrow 5s^2~^1S_0$ transition are well separated by the hyperfine splitting of $793(8)$~MHz, as seen in Fig.~\ref{fig3:Yttrium_data}(a). Here and throughout the number in () denotes the statistical standard error; while we do not treat the systematic error, we estimate it to be roughly 20~MHz, dominated by wavemeter calibration.
For the $5s5p~^3P_0^o \leftrightarrow 4d5s~^3D_1$ transition the two hyperfine lines $F'=\frac{1}{2} \leftrightarrow F''=\frac{3}{2}, \frac{1}{2}$ overlap, as shown in Fig.~\ref{fig3:Yttrium_data}(b), but the data is reproduced by the Voigt fit with a $4d5s~^3D_1$ hyperfine splitting of 345(4)~MHz. 
In Fig.~\ref{fig3:Yttrium_data}(c), the spectra of $5s5p~^3P_1^o \leftrightarrow 4d5s~^3D_2$ transition results from three hyperfine lines, with the two strongest showing substantial overlap.
Based on expected transition strengths, the largest feature is assigned to the $F'=\frac{5}{2}\leftrightarrow F''=\frac{3}{2}$ transition, while the $F'=\frac{3}{2}\leftrightarrow F''=\frac{3}{2}$ and $F'=\frac{3}{2}\leftrightarrow F''=\frac{1}{2}$ transitions were determined by their relative position and signal strength. 
The most complicated spectrum, resulting from the $5s5p~^3P_1^o \leftrightarrow 4d5s~^3D_1 $, transition is presented in Fig.~\ref{fig3:Yttrium_data}(d). 
Here, a wide range of splittings can be observed, and their deconvolution relies heavily on the precision of the fitting. 
This fitting identifies the four transitions as $F'=\frac{3}{2}\leftrightarrow F''=\frac{3}{2}$, $F'=\frac{1}{2}\leftrightarrow F''=\frac{3}{2}$, $F'=\frac{3}{2}\leftrightarrow F''=\frac{1}{2}$, and $F'=\frac{1}{2}\leftrightarrow F''=\frac{1}{2}$ in order of increasing frequency.
While we are unaware of a previous measurement of the hyperfine constant of the $5s5p^3P_1$ level, the determined $4d5s ^3D_1$ and $4d5s ^3D_2$ hyperfine constants agree well with the previous measurement from W{\"a}nnstr{\"o}m et al.~\cite{wnnstrm_high_1994}.  
\begin{table}[h!]
\begin{tabular}{ccccc}
\hline\hline
\multicolumn{2}{c}{Atomic levels} & \multicolumn{2}{c}{Hyperfine levels} & Frequencies \\
\cline{1-4}
\text{Lower~} &  \text{~Upper}&\text{Lower~}($F''$) &  \text{~Upper} ($F'$)&(GHz)\\
\hline
$5s^2~^1S_0$ &  $5s5p ~^3P_1$ & $1/2$ &  $1/2$ & 712793.583(5) \\
 && $1/2$&$3/2$ & 712794.376(7) \\
\hline
$4d5s ~^3D_1$ &  $5s5p ~^3P_0$ & $3/2$ & $1/2$ & 677676.770(2) \\
 && $1/2$ &$1/2$ & 677677.103(3) \\
\hline
$4d5s ~^3D_2$&$5s5p ~^3P_1$ & $3/2$&$3/2$ & 681462.714(12) \\
 && $5/2 $&$3/2$ & 681463.251(4) \\
 &&$3/2$&$1/2 $& 681463.501(5) \\
 \hline
$4d5s ~^3D_1 $&$5s5p ~^3P_1$ & $3/2$&$3/2$ & 687605.032(3) \\
 && $1/2$&$3/2$ & 687605.392(15) \\
 && $3/2$&$1/2$ & 687605.841(12) \\
 && $1/2$&$1/2$ & 687606.202(7)\\
 \hline\hline
\end{tabular}
\caption{The frequencies of the hyperfine-resolved transitions measured in this work. Error bars are statistical. Systematic error is estimated as roughly 20~MHz.}
\label{Tab1:HfFrequency}
\end{table}

\begin{table}[h!]
    \centering
    \begin{tabular}{rrrr}
    \hline\hline
       Levels  & $A$ (expt) (MHz) & $A$ (theory) (MHz) & Ref. \cite{wnnstrm_high_1994} (MHz) \\
       \hline
        $5s5p\,\,^3P_1$ &-532(7) & -557(15) & - \\
        $4d5s\,\, ^3D_1$ &225(12) & 234(7) & 232.2(1.3)\\
        $4d5s\,\, ^3D_2$ &-215(5) & -224(10) & -222.9(8)\\
        \hline\hline
    \end{tabular}
    \caption{The comparison of the hyperfine $A$ coefficients measured and calculated in this work (expt, theory) and work from W{\"a}nnstr{\"o}m et. al.\cite{wnnstrm_high_1994}. 
    A complete listing of hyperfine coefficients calculated in this work is found in Table.~\ref{final_table}.
    }
    \label{tab:hfconst}
\end{table}

\section{Electronic structure calculations}
\label{sec:electronic_structure}
To contextualize the experimental results and provide information on many of the unmeasured features of the $\Y$ spectrum, we performed high-level electronic structure calculations.
The results presented here were achieved using a powerful method for atomic structure calculation based on a hybrid of configuration interaction (CI) and linearized coupled-cluster (CC) techniques \cite{CI+All-order}. 
This approach applies the very precise but resource consuming CI method only for the two valence electrons of Y$^+$. 
The influence of other (core) electrons, such as core-core and core-valence correlations, is taken into account by constructing an effective Hamiltonian with a CC techniques. 
We refer to this approach as the CI+all-order method. 
To estimate the accuracy of this method we duplicated all calculations with an effective Hamiltonian built from second-order many-body perturbation theory (MBPT); we refer to this method as CI-MBPT. 
The difference in the results obtained with more precise CI+all-order method and CI-MBPT is used to estimate the uncertainties of our calculations $\Delta_1$. 

\subsection{Calculated transition rates, lifetimes, and hyperfine constants}
For singly-ionized yttrium with a nuclear spin $I = 1/2$ and two valence electrons, we focused on the group of states and transitions between all low-lying levels from the ground state $5s^2 {}^1S_0$ up to $4d5p {}^3D_3$ with an energy of 29214.0~cm$^{-1}$, i.e. the states shown in Fig.~\ref{fig:level_diagram_combo}(a). 
The resulting state energies, hyperfine constants, and lifetimes are shown in Tab.~\ref{final_table}.

Reduced matrix elements for transitions between the first fourteen even states and the group of odd states were calculated.  
For electric dipole (E1) transitions several corrections to the E1 operator beyond the random-phase approximation (RPA) were included: the core-Brueckner ($\sigma$), structural radiation (SR), two-particle (2P), and normalization (Norm) corrections \cite{hyp_structure,pol_ytt, hyp_ytt}.  
The resulting data are shown in Tab.~\ref{trans_tab} (see Appendix). 
In the cases where $\sigma$-SR-2P-Norm corrections were used,
the resulting error estimation $\Delta$ is defined as follows: 
\begin{equation}
\Delta=\sqrt{\sum_i{(\delta_i/2)^2}+\Delta_1^2} \ ,
\end{equation}
where $\delta_i$ is the respective correction mentioned above.
Tab.~\ref{trans_tab} also includes the corresponding transition rates $A_{ki}$ in $s^{-1}$, evaluated according to the following expressions for the $E1$, $M1$, and $E2$ transitions \cite{NIST_ASD}: 
\begin{equation}
A_{ki}=\frac{2.0261 \times 10^{15}}{(2J_k+1)\lambda^3}|\braket{i||E1||k}|^2 
\end{equation} 
\begin{equation}
A_{ki}=\frac{2.697 \times 10^{10}}{(2J_k+1)\lambda^3}|\braket{i||M1||k}|^2 
\end{equation} 
\begin{equation}
A_{ki}=\frac{1.1199 \times 10^{13}}{(2J_k+1)\lambda^5}|\braket{i||E2||k}|^2 ,
\end{equation}
where the indexes $i$ and $k$ correspond to the lower and upper levels, respectively, $J$ is the total electronic angular momentum, $\lambda$ is the transition wavelength in $\textrm{nm}$, and $\braket{i||E1||k}$, $\braket{i||M1||k}$, $\braket{i||E2||k}$ are electronic dipole, magnetic dipole, electric quadruple reduced matrix elements, respectively. Here, $\braket{i||E1||k}$ and  $\braket{i||E2||k}$ are expressed in atomic units, and $\braket{i||M1||k}$ is given in Bohr magnetons.\footnote{We note that in Hartree atomic unit system the atomic unit for $\braket{i||M1||k}$ is  2$\mu_B$, but the M1 matrix element is usually expressed in $\mu_B$.}

From this data, the lifetime $\tau$ of the states are calculated according to:
\begin{equation}
\tau_i=\frac{1}{\sum_{k<i}^{}A_{ik}}
\end{equation}
and presented in Tab.~\ref{final_table} for the first 25 levels of $\mathrm{Y}^+$.

Finally, the calculated hyperfine constants, including the RPA and $\sigma$-SR-2P-Norm corrections, are also shown in Tab.~\ref{final_table}.

\begin{table}                                                     
\caption{  Experimental $E_{ex}$ \cite{NIST_ASD} and calculated $E_{th}$ energy levels (in cm$^{-1}$), hyperfine constants A$_{hfs}$ (in MHz) and life times $\tau$ (in s) for the first 25 levels of $\Y$.}                                                   
\label{final_table}     
  \begin{ruledtabular}    
  \begin{tabular}{llrrcc}    
Conf & Term & $E_{ex}$& $E_{th}$ & A$_{hfs}$ & $\tau$ \\                                      
\hline
5s$^2$                  &a $^1$S$_0$& 0  &0& 0&$\infty$\\            
4d5s                    &a $^3$D$_1$& 840.2  &967.6& 234(7)&4.5(5)$\times 10^{10}$ \\    
                        &a $^3$D$_2$& 1045.1  &1179.1&-224(10) &4.871(11)$\times 10^{3}$ \\    
                        &a $^3$D$_3$& 1449.8  &1591.8& -241(7)&853.5(2.0) \\       
4d5s                    &a $^1$D$_2$& 3296.2  &3429.5& 24(5)& 99(8)\\     
4d$^2$                  &a $^3$F$_2$& 8003.1  &7993.9& -101.2(2.4)&14.07(24)\\     
                        &a $^3$F$_3$& 8328.0  &8325.8& -26.5(9)&12.29(21) \\     
                        &a $^3$F$_4$& 8743.3  &8752.4& 3.7(5)& 11.35(21)\\     
4d$^2$                  &a $^3$P$_0$& 13883.4  &13986& 0& 6.98(12)$\times 10^{-1}$\\     
                        &a $^3$P$_1$& 14018.3  &14119.3& 58.6(1.1)&6.76(8)$\times 10^{-1}$\\     
                        &a $^3$P$_2$& 14098.1  &14233.0& 27.5(1.1)&7.10(11)$\times 10^{-1}$\\  
4d$^2$                  &b $^1$D$_2$& 14832.9  &15097.6& -30.5(5)&9.7(3)$\times 10^{-1}$ \\     
4d$^2$                  &a $^1$G$_4$& 15682.9  &15863.7& -40.9(1.1)& 1.36(6)\\    
5s5p                    &z $^3$P$^o_0$& 23445.1  &23957.5&0 & 5.4(4)$\times 10^{-8}$\\    
                        &z $^3$P$^o_1$& 23776.2  &24283.7&-557(15) &5.1(4)$\times 10^{-8}$\\    
                        &z $^3$P$^o_2$& 24647.1  &25167.0 & -368(6)&5.6(1.0)$\times 10^{-8}$ \\ 
4d$^2$                  &$^1$S$_0$& 25070.3  &25753.4& 0& 1.29(20)$\times 10^{-2}$\\     
4d5p                    &z $^1$D$^o_2$& 26147.3  &26440.9 & -111(10)&6.71(25)$\times 10^{-9}$\\        
4d5p                    &z $^3$F$^o_2$& 27227.0  &27596.9& -87.5(2.9)&6.39(12)$\times 10^{-9}$\\     
                        &z $^3$F$^o_3$& 27532.3 &27941.4& -64.0(1.4) &6.41(13)$\times 10^{-9}$ \\     
                        &z $^3$F$^o_4$& 28394.2  &28814.5& -26.0(7)&6.04(13)$\times 10^{-9}$\\     
4d5p                    &z $^1$P$^o_1$& 27516.7  &27844.9& 27(9)& 6.3(1.1)$\times 10^{-9}$\\   
4d5p                    &z $^3$D$^o_1$& 28595.3  &28885.0& -85(16)& 4.7(6)$\times 10^{-9}$\\     
                        &z $^3$D$^o_2$& 28730.0  &29019.5& -25.6(8)& 4.45(7)$\times 10^{-9}$\\     
                        &z $^3$D$^o_3$& 29214.0  &29516.7& -21.5(1.0)& 4.38(9)$\times 10^{-9}$\\          

\end{tabular}                                                                   
\end{ruledtabular}

\end{table}

\subsection{Hyperfine quenching of the $\spThreePZero$ level}
\label{sec:quench}
The transition between the $\spThreePZero$ and $\ssOneSZero$ states of Y$^+$ is strictly forbidden for all electromagnetic multipole operators.
In principle, this means that operations on ions in the metastable states, such a laser cooling via the $\spThreePOne \leftrightarrow \dsThreeDOne$ transition, will not lead to population in the ground $\ssOneSZero$ state.
This attractive feature provides isolation of qubits stored in the ground-state manifold from operations performed on qubits stored in metastable states.
However, hyperfine-induced mixing between the $\spThreePZero$ and $\spThreePOne$ states provides a non-zero decay moment, potentially complicating this picture. 
We estimate the $\spThreePOne \rightarrow \ssOneSZero$ decay following Ref.~\cite{Porsev_Derevianko} as:
\begin{equation}
    A_{\mathrm{HFI}}\left(n s n p\,{ }^{3}P_{J} ; F \rightarrow  n s^{2}\,{ }^{1}S_{0}\right)
    =\frac{4 \alpha^{3}}{9} \omega_{J}^{3}\left|\sum_{k} S_{k}\right|^{2}
\label{quench_trans}
\end{equation}
where $a \approx 1/137$ is the fine-structure constant, $\omega_J=E\left(n s n p\,{}^{3}P_{J}\right)- E\left(n s^{2}\,{ }^{1} S_{0}\right)$ is the transition frequency, and $S_k$ is defined as follows:
\begin{equation}
    \begin{aligned}
S_{k}=  \left\langle I\left\|\mathcal{M}^{(k)}\right\| I\right\rangle \sum_{\gamma^{\prime}, J^{\prime}}\left\{\begin{array}{lll}
I & I & k \\
J & J^{\prime} & F
\end{array}\right\} \times \\ \times \frac{\left\langle n s^{2}\,{ }^{1}S_{0}\|D\| \gamma^{\prime} J^{\prime}\right\rangle\left\langle\gamma^{\prime} J^{\prime}\left\|\mathcal{T}^{(k)}\right\| n s n p\,{}^{3}P_{J}\right\rangle}{E\left(\gamma^{\prime} J^{\prime}\right)-E\left(n s n p\,{}^{3}P_{J}\right)} .
\end{aligned}
\label{Sk}
\end{equation}
Here, the nuclear and electronic angular momenta, $I$ and $J$, are conventionally coupled to form a state with total angular momentum $F$ and projection $M_F$, while $\gamma$ denotes all other relevant atomic quantum numbers. The operator $D$ is the electric dipole operator, $\mathcal{M}^{(k)}$ are the nuclear multipole moments of rank $k$, and $\mathcal{T}^{(k)}$ are the associated electronic operators of the same rank (with odd $k$ corresponding to magnetic moments and even $k$ to electric moments).
The $\mathcal{M}^{(k)}$ and $\mathcal{T}^{(k)}$ appearing in (\ref{Sk}) arise from the Hamiltonian of the hyperfine interaction between the nucleus and atomic electrons, which can be written as
\begin{equation}
    H_{\mathrm{HFI}}=\sum_{k}\left(\mathcal{M}^{(k)}\cdot \mathcal{T}^{(k)}\right).
\label{H_HFI}
\end{equation}
The scalar product of spherical tensors
\[
    \mathcal{M}^{(k)}\cdot \mathcal{T}^{(k)}=\sum_{q=-k}^{k}(-1)^q \ \mathcal{M}_q^{(k)}\mathcal{T}_{-q}^{(k)}
\]
is expressed in terms of spherical harmonics $C_q^{(k)}({\mathbf{\hat r}})$ in Racah normalization:
\begin{eqnarray}
\begin{aligned}
    \mathcal{M}^{(k)}_q&=\sum_{i\in nucleus} e_ir_i^{k}C_q^{(k)}(\mathbf{\hat{r}}), \\
    \mathcal{T}^{(k)}_q&=\sum_{i\in e} \frac{C_q^{(k)}(\mathbf{\hat{r}})}{r_i^{k+1}}.
\end{aligned}
\label{M,T}
\end{eqnarray}
The above expression (\ref{Sk}) for $S_{k}$ follows from first-order perturbation theory with $H_{\mathrm{HFI}}$, Eq.~(\ref{H_HFI}), treated as a perturbation.

In the present case $J^{\prime}=1$ due to the electric-dipole selection rules and only the magnetic $(k=1)$ HFI coupling causes the E1 transitions. 
Given the nuclear spin is $I = 1/2$, the hyperfine interaction is given by the magnetic dipole term and the decay rate of Eq.~\eqref{quench_trans} simplifies to 
\begin{equation}
\begin{aligned}
    A_{\mathrm{HFI}}\left(5s5p \ { }^{3} P_{0} ; F=I=\frac{1}{2} \rightarrow 5s^2 \ { }^{1} S_{0}\right)= \\ =\frac{4 \alpha^{3}}{9} \omega_{0}^{3} \mu^{2}\times \left|\left\langle\delta \Psi\left\|\mathcal{T}_{\text {eff }}^{(1)}\right\| 5s5p \ ^{3} P_{0}\right\rangle\right|^2 ,
\end{aligned}
\label{tran_rate}
\end{equation}
where the transition frequency is $\omega_{0}= E_{(^3P_0)}-E_{(^1S_0)}$, the nuclear moment is \[ \mu=\braket{IM_I|\mathcal{M}^{(1)}|IM_I}=-0.137298 \] in nuclear magnetons \cite{Stone2019}, and
\begin{equation}
\begin{aligned}
   &\left\langle\delta \Psi\left\|\mathcal{T}_{\text {eff }}^{(1)}\right\| 5s5p \ ^{3} P_{0}\right\rangle = \\ &\sum_{\gamma^{\prime}} \frac{\left\langle 5s^2 \ { }^{1} S_{0}\|D\| \gamma^{\prime} J^{\prime}=1\right\rangle\left\langle\gamma^{\prime} J^{\prime}=1\left\|\mathcal{T}^{(1)}\right\| 5s5p \ { }^{3} P_{0}\right\rangle}{E\left(\gamma^{\prime} J^{\prime}=1\right)-E\left(5s5p \ { }^{3} P_{0}\right)}.
\end{aligned}
\label{tau_eff}
\end{equation}
This expression (\ref{tau_eff}) can be computed using the
Sternheimer-Dalgarno-Lewis method \cite{Sternheimer, Dalgarno_Lewis}, where $\left\langle\delta \Psi\right|$ satisfies the inhomogeneous Schrodinger equation:
\begin{equation}
    \left[H_{\mathrm{eff}}-E\left(5s5p \ ^{3} P_{0}\right)\right]|\delta \Psi\rangle=D_{\mathrm{eff}}\left|5s^{2} \ { }^{1} S_{0}\right\rangle
\end{equation}
Hence, we obtained 
\[
  \left\langle\delta \Psi\left\|\mathcal{T}_{\text {eff }}^{(1)}\right\|5s5p \ { }^{3} P_{0}\right\rangle = 1.26 \cdot 10^6 \ \rm{MHz} 
\label{T}
\]
and
\[
    A_{\mathrm{HFI}}\left(5 s 5 p\ { }^{3} P_{0} ; F \rightarrow 5s^{2}\ { }^{1} S_{0}\right)=6.0 \cdot 10^{-3} \ \ \s^{-1}.
\label{A}
\]
Given the calculated lifetime of the $5s5p~^3P_1^o$ state, this yields a branching fraction of $\sim 3\times 10^{-10}$ for hyperfine-induced decays into the ground state manifold. Hyperfine-induced decays into other manifolds, along with non-dipole decays, are shown in Table.~\ref{tab:lasercooling}.

\section{Quantum information processing}\label{sec:quantum_ops}
The rich structure that comes with an additional valence electron provides numerous opportunities and modalities of operation for quantum logic with $\Y$.
Therefore, in this section we focus only on several broad classes of methods that should be of generic use.
One such attractive modality is \emph{omg}-style operation where information is stored in the ground state, as a nuclear qubit with low-field sensitivity, and coherently shuttled into a metastable state within the $4d5s$ manifold for gate operations.
These metastable states allow gates to be performed via Raman laser operation using excited states in the $5s5p$ or $4d^2$ manifold, as well as via magnetic dipole couplings to microwave- or radio-frequency fields. 
It is shown that $\Y$ is particularly attractive for magnetic field gradient gates as it provides both magnetic-field-insensitive storage in the nuclear qubit and magnetic-field-sensitive metastable states, desirable for gate operation, in one species.

Another modality involves storing the qubit in a pair of hyperfine clock states within a long-lived metastable manifold. 
In this case gate operations could be done directly on the hyperfine qubit, and one or both qubit states could potentially be mapped into a different manifold for low-crosstalk measurement operations. 
Here we focus on the $\dsThreeDOne$ manifold, although other attractive possibilities appear to exist.

Therefore, in what follows we use the spectroscopy and calculations described above to estimate the efficiency and error rates of the processes necessary for quantum logic with $\Y$ qubits, including laser cooling, state preparation and measurement, qubit storage, coherent shuttling, and gates. In particular, we illustrate the possibility of using the nuclear spin qubit or a metastable clock qubit for information storage, while performing spectrally isolated operations in other manifolds.
The discussion is not exhaustive, but illustrates the wide range of possibilities for information processing with this ion. 

\subsection{Laser cooling}
While sympathetic cooling with e.g. laser-cooled $\mathrm{^{88}Sr}^+$ can be used to efficiently cool $\Y$, it is helpful to identify a cycling transition for direct cooling. 
Laser cooling on the $\spThreePZero \leftarrow \dsThreeDOne$ at 442~nm appears to be an attractive choice.
First, the decay rate of the $\spThreePZero$ is calculated as $\Gamma = 2\pi \cdot 2.95(24)~\MHz$, leading to a Doppler temperature limit of $\approx 70~\mu$K, meaning little to no resolved sideband cooling will be required for preparation of ground-state cooled samples.
Second, owing to selection rules, the transition is nearly closed with a calculated out-of-cycle branching ratio of $\approx 5\%$ to $\ddThreePOne$, which can be repumped with a laser at 1061~nm. $\ddThreePOne$ also has M1 and E2 decay channels, which result in a calculated lifetime of $0.676(8)\,\s$, which should be sufficiently long to allow Doppler cooling with minimal leakage into other metastable manifolds. 
If necessary, leaked ions can be identified by checking for fluorescence on this transition (Sec.~\ref{sec:measurement}) prior to initialization. Alternatively, cooled ions can be deterministically flushed into the ground-state manifold, at the cost of a large number of low-power lasers to repump the metastable manifolds.

This transition could also be used for sympathetically cooling $\Y$ ions hosting qubits stored in their ground-state manifolds, with periodic leakage detection or flushing of coolant ions, thereby removing the need for a distinct coolant ion species.

Due to hyperfine mixing, $\spThreePZero$ can decay to other states, such as $\ssOneSZero$.
Using the method described in Sec.~\ref{sec:quench}, we have calculated these hyperfine-mixing-enabled decay rates, which are shown in Tab.~\ref{tab:lasercooling}, and find they occur with a total branching ratio of $\approx 2\times10^{-9}$.
The leading decay is an M2 decay to the $\dsThreeDTwo$ at 446~nm. If population in this state is also re-pumped, the out-of-cycle branching ratio decreases by roughly a factor of three.
Additional repumping can be used to further decrease the leakage, however, as the next leading decay is to the $5s^2~^1S_0$, this may be undesirable if the nuclear qubit is used for storage. 
Regardless, these branching ratios should allow scattering of more than enough photons for laser cooling. 

If traditional sideband cooling is necessary, it could be performed on several of the narrow lines, such as the $4d^2~^3P_J \leftrightarrow \dsThreeDOne$ or $4d^2~^3F_J \leftrightarrow \dsThreeDOne$ transitions, or by using a Raman transition.

\begin{table}                                                     
\caption{  Comparison of the hyperfine quenching rates with the non-dipole rates for the $5s5p~^3P_0^o$ state.}                                                   
\label{tab:lasercooling}    
\setlength{\extrarowheight}{3.5pt}                                              
\begin{ruledtabular}                                                          
  \begin{tabular}{llcccc}
Config.     & Term & WL, nm & \multicolumn{3}{c}{Tr. Rate, s$^{-1}$} \\
&&&Non-dipole & Quenching & Final \\ 
\hline
        $5s^2$ & $^1S_0$ & 426.53 & 0.00     & 6.01E-03 & 6.01E-03 \\
        $4d5s$ & $^3D_2$ & 446.43 & 1.01E-05 & 2.34E-02 & 2.35E-02 \\
               & $^3D_3$ & 454.64 & 1.93E-06 &          & 1.93E-06 \\
               & $^1D_2$ & 496.31 & 1.31E-04 & 3.60E-04 & 4.91E-04 \\
        $4d^2$ & $^3F_2$ & 647.59 & 1.12E-06 & 1.58E-04 & 1.59E-04 \\
               & $^3F_3$ & 661.51 & 1.04E-08 &          & 1.04E-08 \\
               & $^3P_0$ & 1045.84& 0.00     & 7.61E-04& 7.61E-04 \\
               & $^3P_2$ & 1069.86& 1.75E-06 & 1.01E-03 & 1.01E-03 \\
               & $^1D_2$ & 1161.14& 9.75E-07 & 2.00E-05 & 2.10E-05 \\
    \end{tabular}
\end{ruledtabular} 
\end{table}

\subsection{Qubit storage}
\subsubsection{Nuclear spin qubit}The $\ssOneSZero$ ground-state manifold provides a natural nuclear spin storage qubit which offers favorable sensitivity to environmental magnetic fields compared to typical trapped-ion qubits. 
In particular, while hyperfine clock qubits acquire qubit frequency shifts due to the presence of RF trapping potentials~\cite{Gan2018}, a nuclear spin qubit acquires zero net shift for RF magnetic fields parallel to the quantization axis due to the linear dependence of the states on the external magnetic field; and for perpendicular fields, it is highly suppressed by the small nuclear magnetic dipole moment and the large frequency difference between typical RF trapping frequencies ($\gtrsim 20\,\MHz$) and the nuclear Zeeman splitting ($\lesssim 100\,\kHz)$. 
Thus, the magnetic field sensitivity of a nuclear qubit compares favorably to a typical $M = 0$ (`zero field') hyperfine clock qubit, which must be used with a finite bias magnetic field to remove the degeneracy with the spectator Zeeman states (with $M\ne 0)$. 
This finite field causes the qubit to acquire a linear field sensitivity. 
For instance, at a bias field of $B = 4$~G, the zero-field clock qubit in $^{171}\textrm{Yb}^+$ defined by the $S_{1/2}, F=(0,1),\; M=0$ states has a linear field sensitivity of $2.5\,\kHz/\Gauss$, while the $^{89}\textrm{Y}^+$ nuclear spin qubit has a sensitivity of $0.21\,\kHz/\Gauss$ (independent of bias field).

The weak coupling of the nuclear spin qubit to external fields, and the small frequency splitting between the qubit states, however, make it more challenging to perform quantum gates directly on the qubit. 
Both magnetic dipole and Raman electric dipole (E1) transitions (when the beams are far-detuned from the intermediate excited-state manifold) between the states are highly suppressed. 
One option for a Raman single-qubit gate is to use a relatively small detuning from the excited-state manifold and a sufficiently long gate time at a large bias magnetic field, as demonstrated in neutral ${}^{171}\mathrm{Yb}$~\cite{Muniz2025}. This will come at the cost of additional spontaneous emission errors due to the small detuning needed to couple the nuclear spin states. Moreover, achieving the same nuclear spin splitting in $\Y$ would require $\approx 3.6\times$ the magnetic field for as for ${}^{171}\mathrm{Yb}$, owing to the smaller nuclear magnetic moment. Nuclear spin states have also been addressed in neutral $^{87}\mathrm{Sr}$ by applying an AC Stark shift with an additional beam~\cite{Barnes2022}, though this introduces sensitivity to the intensity fluctuations of that beam.

Another option is to coherently shelve and de-shelve between a metastable manifold with hyperfine structure to perform operations. Shelving is favorable for suppressing crosstalk in large-scale systems, as it leaves stored qubits spectrally isolated from electromagnetic fields applied for gates and measurement, at the cost of introducing additional technical complication and possibly additional errors from the shelving/de-shelving process. This is the method we consider, and is discussed in Sec.~\ref{sec:shelving}.

\begin{figure}
 \includegraphics[width=1.0\columnwidth]{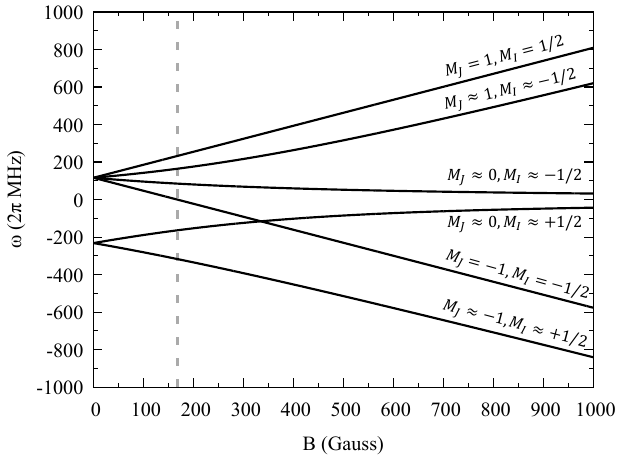}
 \caption{\label{fig:3D1_energies}Energies of the $\dsThreeDOne$ manifold vs. magnetic field, assuming hyperfine coefficient $A=232.2\,\MHz$~\cite{wnnstrm_high_1994}. Vertical dashed line shows the hyperfine clock point $B\approx 168\,\Gauss$ for the states $\ket{F=3/2,M=1/2}$, $\ket{F=1/2,M=1/2}$.}
\end{figure}

\subsubsection{Metastable hyperfine clock qubit}~The long-lived metastable manifolds offer alternative qubit storage modalities. In particular, at a magnetic field of $B \approx 168\,\Gauss$, the $\dsThreeDOne$ manifold hosts a finite-field hyperfine clock qubit between hyperfine states that correlate to the $\ket{F = 1/2, M=1/2}$ and  $\ket{F = 3/2, M=1/2}$ states with a qubit frequency of $\approx 330\,\MHz$ and a quadratic sensitivity of $\approx 0.7\,\textrm{kHz}/\textrm{G}^2$.
This provides a convenient trapped-ion clock qubit which could be used with either laser-, microwave-, or radiowave-based gates, with a much smaller qubit frequency splitting than typical zero-field hyperfine qubits (such as the $\approx 12.6\,\GHz$ zero-field qubit in $\Yb$). 
To mitigate crosstalk from measurement, one may optionally shelve to a different manifold for this operation; see Sec.~\ref{sec:measurement}.

Other metastable manifolds may also offer attractive qubit storage, but we leave their detailed analysis to future studies.

\subsection{Nuclear spin shelving} \label{sec:shelving}

In this section we present several coherent shelving methods which enable simultaneous mapping of the nuclear spin qubit to a metastable manifold, circumventing potential bit-flip errors due to the small Zeeman splitting of the nuclear spin. We discuss methods using Raman beams and analyze errors due to coherent off-resonant coupling to neighboring states, spontaneous emission errors, and errors from polarization imperfections of the beam. We also briefly discuss methods for optical shelving. Simultaneous shelving of both qubit states is robust to phase drifts of the laser system which vary slowly compared to the mapping pulse time, as both qubit states acquire the same, and therefore unobservable, phase error~\cite{Yang2022}.

\subsubsection{$M=\pm 1/2$ circularly-polarized Raman shelving} 
\begin{figure}
 \includegraphics[width=0.9\columnwidth]{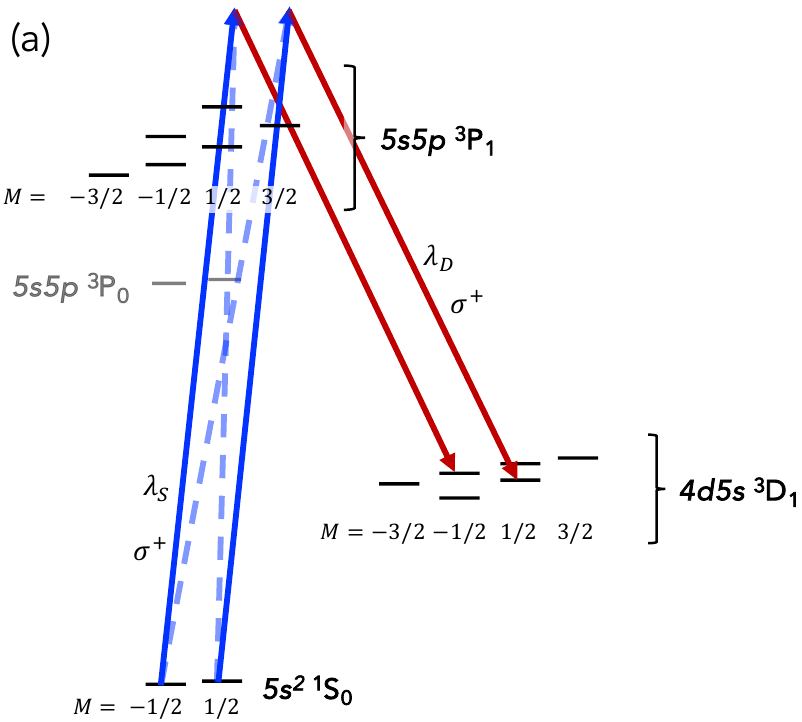}
 \includegraphics[width=0.9\columnwidth]{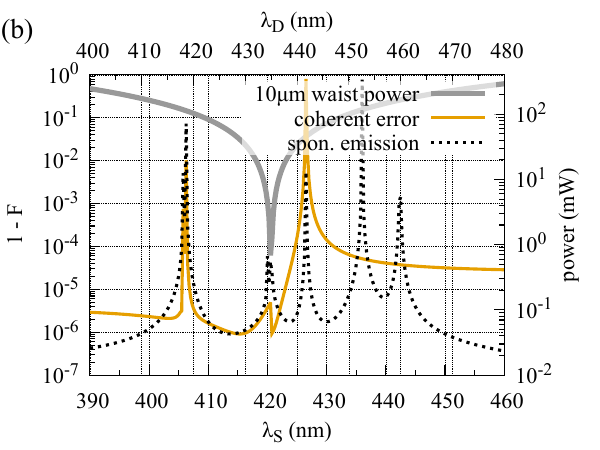}
 \includegraphics[width=0.9\columnwidth]{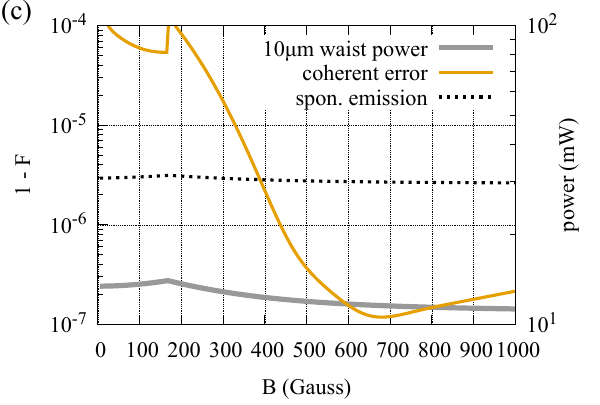}
 \caption{(a) Transitions for coherent Raman shelving of the nuclear spin qubit to the long-lived metastable $\dsThreeDOne$ manifold using circularly polarized light. Dashed lines show unwanted near-resonant spin-flip transitions which are suppressed at larger magnetic fields when the single-photon detuning from the excited-state manifold is large compared to the hyperfine splitting. (b) Estimated errors from spontaneous photon scattering and polarization errors for the coherent shelving scheme of (a) as a function of wavelength, at $B=400\Gauss$. A pointing error of $1$ degree and a polarization error (due to rotation of the polarization vector around the $\tmmathbf{k}$-vector) of $1\%$ are included. (c) Estimated errors from spontaneous photon scattering and polarization errors for the coherent shelving scheme of (a) at $\lambda=419\nm$ (chosen to balance spontaneous emission error and the laser power requirement), as a function of bias magnetic field strength $B$. Pointing and polarization errors are the same as those in (b).}.
 \label{fig:circular_shelving_combo}
\end{figure}

In this method, a pair of phase-locked, co-propagating, circularly polarized beams target the transitions $\ssOneSZero \leftrightarrow \spThreePOne$ and $\spThreePOne \leftrightarrow \dsThreeDOne$ in order to map the nuclear spin qubit states into the $\dsThreeDOne$ manifold via Raman transitions. The qubit states can either be mapped individually or simultaneously. 
Circular polarization is used to avoid cancellations that suppress the coupling for linear polarizations. Fig.~\ref{fig:circular_shelving_combo}a shows the transition diagram for this scheme at $B \approx 400$~G.

In the presence of nonzero pointing error, a circularly polarized beam acquires a $\pi$-polarization component, which can drive unwanted transitions between nuclear spin states. 
However, for large bias fields, the $\dsThreeDOne$ states acquire increasingly well-defined values of the nuclear spin, $\langle I_z \rangle \approx \pm \; \frac{1}{2}$ (Fig.~\ref{fig:3D1_energies}). 
When the single-photon detunings of the beams are large compared to the hyperfine splitting of the excited-state manifold, accidental spin flips due to polarization errors are suppressed.
Therefore, at each bias magnetic field value, the target shelving states for $\ket{\downarrow}$ and $\ket{\uparrow}$ in $\dsThreeDOne$ are chosen to be those with nuclear spin expectations $\langle \hat{I}_z\rangle$ closest to $-1/2$ and $+1/2$, respectively. 

Fig.~\ref{fig:circular_shelving_combo}b and Fig.~\ref{fig:circular_shelving_combo}c show the calculated shelving infidelity of this scheme as a function of Raman laser wavelength and bias magnetic field, respectively. The calculations were done with a shelving pulse time of $T=20\,\mu\s$ with $\sin^2$ turn-on/turn-off time of $10\,\mu\s$. A $1\deg$ pointing error and $1\%$ polarization error on the beam were included. "Coherent" errors, including those induced by the polarization and pointing errors, were determined by a wavefunction evolution, which naturally omits spontaneous emission (SE) errors. Calculation of the latter were determined using an effective master equation for Raman scattering, described in Appendix~\ref{appendix:quantum_ops}. The choice of shelving states changes at approximately $B\approx 168~\Gauss$, leading to the discontinuity seen in Fig.~\ref{fig:circular_shelving_combo}c.

\subsubsection{$M=\pm 3/2$ (stretched) linear-polarized Raman shelving} In this method, one or both nuclear spin states are shelved to the $\dsThreeDOne$, $M=\pm 3/2$ (``stretched'') states using linearly polarized beams.
The pure nuclear spin of the stretched states helps protect the qubit from accidental spin flips due to polarization errors, allowing this method to work more accurately at low bias magnetic fields than the $M = \pm 1/2$ method. 
The disadvantage is that it is more challenging to intentionally drive spin flips between the $M = \pm 3/2$ states in the $\dsThreeDOne$ manifold or transfer into the finite field clock qubit states.

\begin{figure}
 \includegraphics[width=0.9\columnwidth]{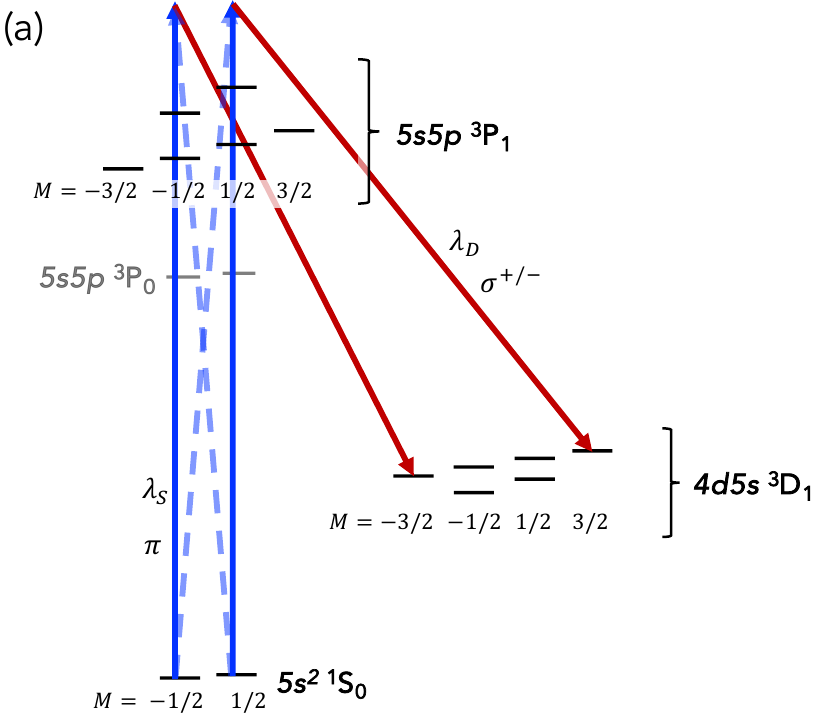}
 \includegraphics[width=0.9\columnwidth]{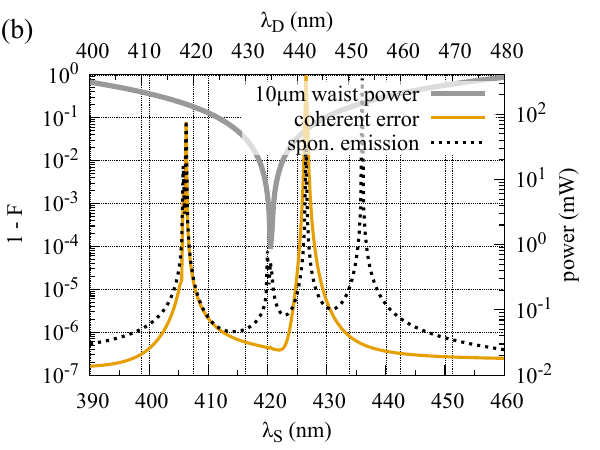}
 \includegraphics[width=0.9\columnwidth]{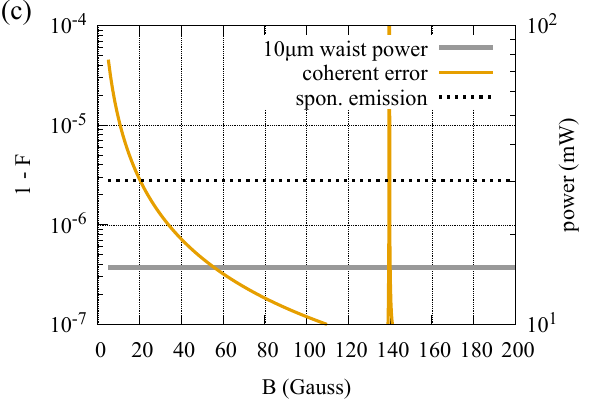}
 \caption{\label{fig:stretched_shelving_combo}(a) Transitions for coherent Raman shelving of the nuclear spin qubit to the $\dsThreeDOne$, $M=\pm 3/2$ states. Dashed lines show unwanted near-resonant spin-flip transitions which are suppressed when the single-photon detuning from the excited-state manifold is large compared to the hyperfine splitting. (b) Estimated errors from spontaneous photon scattering and polarization errors for the coherent shelving scheme of (a) at bias magnetic field strength $B=50$~G. (c) Estimated errors from spontaneous photon scattering and polarization errors for the coherent shelving scheme of (a) at wavelength $\lambda=419~\nm$. Pointing and polarization error are the same as in Fig.~\ref{fig:circular_shelving_combo}. The peak at $B\approx 140\,\Gauss$ is due to an accidental resonance between the applied tones and the $F=3/2,M=-3/2$ $\leftrightarrow$ $F=1/2,M=+1/2$ transition occurring at this field.}
\end{figure}

Fig.~\ref{fig:stretched_shelving_combo}a shows an example transition diagram for the case where one beam is $\pi$-polarized and the other $\sigma^{\pm}$-polarized (polarization vector orthogonal to the bias field); in general, the transition is driven for orthogonal linear polarizations of the two beams. Fig.~\ref{fig:stretched_shelving_combo}b and Fig.~\ref{fig:stretched_shelving_combo}c show the calculated shelving infidelity of this scheme as a function of Raman laser wavelength and bias magnetic field, respectively. As in Fig.~\ref{fig:circular_shelving_combo}, a $1$ degree pointing error and $1\%$ polarization error are included on the beams. Note that mixing of $\dsThreeDOne$ with other metastable manifolds due to hyperfine and magnetic-dipole interactions is not included here; this could reduce the nuclear spin purity of the $M=\pm3/2$ states lead to larger "coherent" errors, but this effect should be small due to the $~6\,\THz$ separation from the nearest metastable manifold.

\subsubsection{Optical shelving}  In this method, one or both nuclear spin states are shelved to a metastable manifold via an optical (single-photon) electric quadrupole transition. 
Table \ref{tab:opticalShelvingTransitions_S} lists the electric quadrupole transitions between $\ssOneSZero$ and metastable manifolds.

\begin{figure}
 \includegraphics[width=0.8\columnwidth]{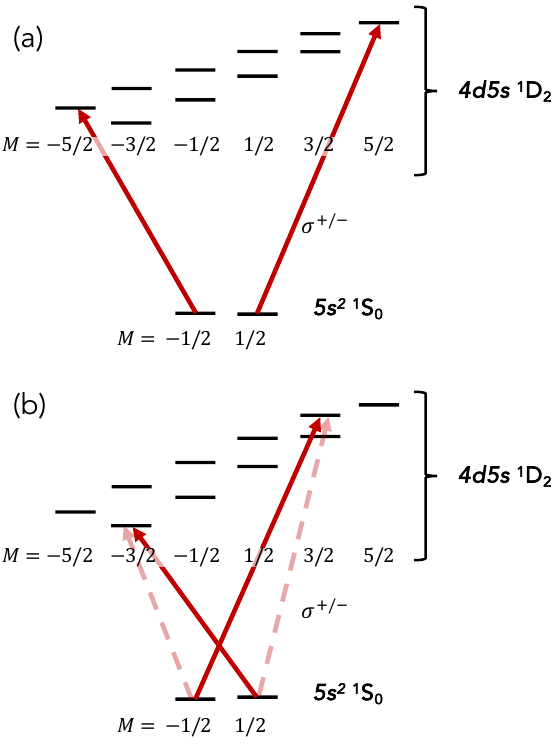}
 \includegraphics[width=0.9\columnwidth]{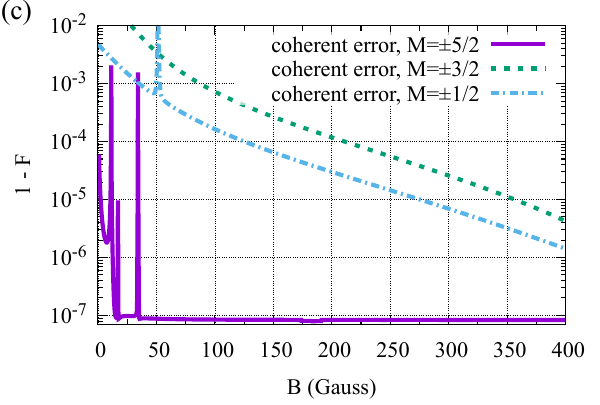}
 \caption{\label{fig:optical_shelving} (a) Transitions for coherent optical shelving of the nuclear spin qubit to the $\dsOneDTwo$, $M=\pm 5/2$ states at $B\approx 50\Gauss$. (b) Transitions for coherent optical shelving of the nuclear spin qubit to $\dsOneDTwo$, $M = \pm 3/2$ states at $B\approx 75\Gauss$ using linearly polarized light. Dashed lines show transitions that are off resonant by the nuclear Zeeman splitting, but are suppressed due to the relative purity of the nuclear spin magnetic projection $I_z \approx 1/2$ of the shelved states. (c) Simulated coherent error for mapping the nuclear spin qubit to $\dsOneDTwo$, $M=\pm 5/2$ (panel (a)), $M=\pm 3/2$ (panel (b)), or $M=\pm 1/2$ states.}
\end{figure}

Shelving via $\Delta M=\pm 2$ transitions to $M=\pm 5/2$ states in $J=2$ manifolds will be the most favorable option for robustness to pointing and polarization errors, as in the $M=\pm 3/2$ Raman shelving technique, since the target states have pure nuclear spin $\langle I_z \rangle = \pm 1/2$. 
This suppresses near-resonant transitions (off-resonant by the nuclear Zeeman splitting) that would flip the spin of the qubit. 
This shelving would enable a two-qubit $ZZ$ gate (see Sec.~\ref{sec:magnetic_gradient_gates}) or state preparation / measurement operations, while gates requiring bit flips (such as general single-qubit gates) would require other techniques due to the difficulty of directly coupling states with $M_I=+1/2$ and $M_I=-1/2$. 
An example is shown in Fig.~\ref{fig:optical_shelving}a for shelving to the $4d5s~^1D_2$ state on the $3.034$~$\mu$m transition. 

Another example is shown in Fig.~\ref{fig:optical_shelving}b, illustrating shelving to an inner pair of states, with $M = \pm 3/2$. 
In this case, as with the circular Raman shelving method, near-resonant spin-flip transitions induced by polarization errors are suppressed by the nuclear Zeeman splitting and the approximate purity of $I_z$ in the metastable states, both of which improve at larger magnetic fields.

Fig.~\ref{fig:optical_shelving}c shows the simulated shelving infidelity for mapping the nuclear spin qubit to $M=\pm5/2$, $M=\pm3/2$, or $M=\pm1/2$ states. A total pulse time of $10\,\us$ was used, including a shaped $\sin^2$ turn-on/turn-off lasting $1\,\us$ each. A $1^\circ$ pointing error and $1\%$ polarization error are included. Decay from the metastable manifold is excluded, so the only error computed is the "coherent" error from off-resonant transitions. The error floor for the $M=\pm 5/2$ scheme is due to far off-resonant transitions allowed by the nominal polarization, while errors for the $M=\pm 3/2$ and $M=\pm1/2$ schemes are dominated by accidental spin flips due to the polarization error, as illustrated in Fig.~\ref{fig:optical_shelving}b.

A variety of strong $E2$ transitions also connect the $\dsThreeDOne$ manifold to higher-energy metastable levels, providing opportunities for operations on a qubit stored in $\dsThreeDOne$. We refer the reader to Appendix~\ref{A} for the available transitions.

\begin{table}
    \centering
    \begin{tabular}{lcr}
    \hline\hline
    Level & $\langle \textrm{level} ||\, E2 \,|| \ssOneSZero\rangle $ & $\lambda (\nm)$ \\
    \hline
        $\dsThreeDTwo$ & $-1.19(6)$ & $9568.7$ \\
        $\dsOneDTwo$ & $-9.49(17)$ & $3033.8$ \\
        $\ddThreeFTwo$ & $-0.407(25)$ & $1249.5$ \\
        $\ddThreePTwo$ & $1.16(3)$ & $ 709.3$ \\
        $\ddOneDTwo$ & $-2.6825(28)$ & $674.2$ \\
        \hline
    \end{tabular}
    \caption{Quadrupole transitions connecting the ground-state manifold $\ssOneSZero$ to metastable manifolds. Matrix elements are in atomic units.}
    \label{tab:opticalShelvingTransitions_S}
\end{table}

\subsection{State preparation}

The nuclear spin qubit could be prepared in an algorithmic way by reusing one of the shelving schemes of Sec.~\ref{sec:shelving}. By including only one tone of the two-tone beam, one can selectively shelve one of the qubit states with an infidelity comparable to that of the parallel shelving operation. Repeatedly shelving and flushing through the $\spThreePOne$ level would therefore allow for high-fidelity state preparation of one of the nuclear spin qubit states, with infidelity limited by that of the single-tone shelving operation. Flushing through $\spThreePOne$ would require a large number of wavelengths, including 436\,nm, 440\,nm, 488\,nm, 634\,nm, 1011\,nm, 1025\,nm, 1033\,nm, and 1118\,nm (see Appendix~\ref{appendix:electronic_structure}), to cover all E1 decay channels of this manifold. 
Small residual population in other manifolds can be handled at the time of readout by leakage detection (see below). A single additional high-fidelity pulse would allow state preparation in a metastable manifold.

An alternative method of preparing a state in a metastable manifold would be to utilize a narrow quadrupole transition into a different metastable manifold, followed by flushing through an excited-state manifold, similar to the narrow-band optical pumping method of Ref.~\cite{An2022}. For example, to prepare a state in $\dsThreeDOne$, population in unwanted states could be transferred from $\dsThreeDOne$ to $\ddThreePOne$ using the quadrupole transition at $759\nm$, and then flushed back to $\dsThreeDOne$ via the excited $\spThreePZero$ level using $1061\nm$ light.

\subsection{Measurement} \label{sec:measurement}

\subsubsection{$\spThreePZero$ measurement}

\begin{figure}
 \includegraphics[width=.7\columnwidth]{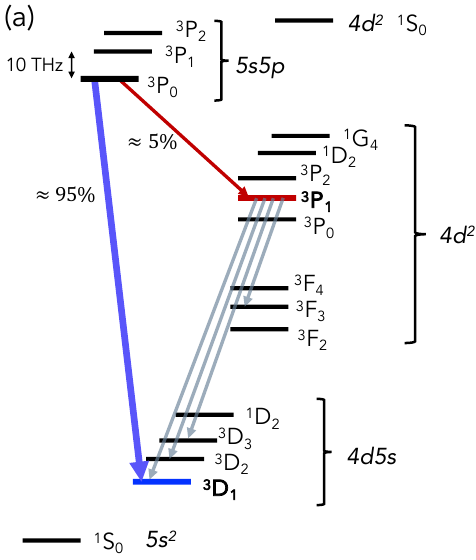}

\vspace{0.5cm}
 
 \includegraphics[width=.7\columnwidth]{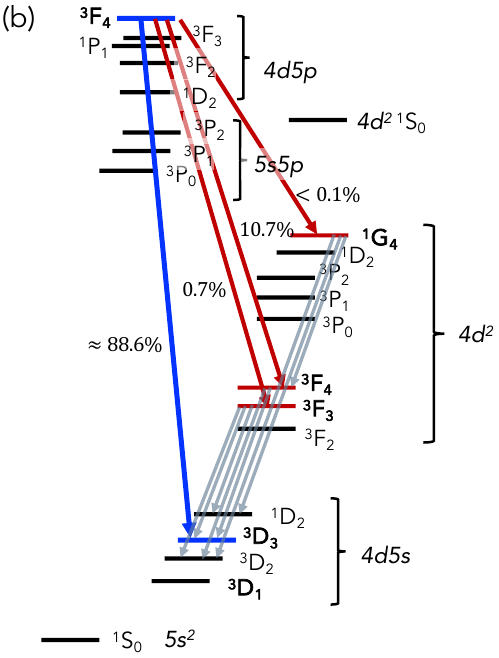}
 \caption{\label{fig:measurement_combo} (a) Decay channels for measurement on the $\spThreePZero \leftrightarrow \dsThreeDOne$ transition. (b) Decay channels for measurement on the $\dpThreeFFour \leftrightarrow \dsThreeDThree$ transition. Percentages shown are calculated branching fractions. Light gray arrows from metastable levels show M1 or E2 decay channels with calculated transition rates $\ge 10^{-2} \,\s^{-1}$. Slower decay channels exist but are not shown.}
\end{figure}

The $\dsThreeDOne$ manifold provides a nearly-closed cycling transition at $442\,\nm$ with the excited $\spThreePZero$ level, allowing for essentially crosstalk-free measurement for nuclear spin qubits stored in the $\ssOneSZero$ manifold. There is one additional weak branching to the relatively long-lived $\ddThreePOne$, which has lifetime $\approx 676\,\ms$ and $50\%$ branching back to $\dsThreeDOne$; see Fig.~\ref{fig:measurement_combo}a. 
Decay of $\spThreePZero$ on the $442\,\nm$ transition occurs at a calculated rate of $2\pi \cdot 2.80(2)\,\MHz$. Assuming a time-averaged population of $0.25$ in the $\spThreePZero$ manifold during measurement, this rate would suggest a measurement time of $~50\,\us$ for the collection of $\sim25$ photons with a $3\%$ detection efficiency, sufficient for high-fidelity bright/dark discrimination when background counts are sufficiently small.

Non-dipole decays and hyperfine quenching of $\spThreePZero$, along with the finite lifetime of $4d^2\,{}^3P_1$, will cause a small but nonzero probability of leakage to other long-lived metastable manifolds during the measurement. The former gives an out-of-cycle branching fraction of $\sim 2\times10^{-9}$ (Table~\ref{tab:lasercooling}). The latter, assuming a time-averaged population of $0.1$ in $4d^2\,{}^3P_1$ during measurement, gives an effective branching fraction of $\sim 0.1 \times 0.714\,\mathrm{s}^{-1} / (2\pi \times 2.8\,\MHz) = 4\times10^{-9}$. Together these imply a leakage probability of $\sim 5\times10^{-6}$ during a $50\,\us$ measurement. Measuring both qubit states successively during readout would allow the detection of leaked qubits (which would register dark for both measurements) and thereby prevent the accumulation of leakage from repeated initialization, gating, and readout operations.

Leaked qubits can either be flushed into the qubit space using a large number of low-power lasers targeting the metastable manifolds, or replaced with freshly loaded and initialized qubits (which have been verified as not leaked during the loading process). Due to the large number of metastable manifolds and decay channels between them, unless all metastable manifolds are periodically flushed, qubits may need to be checked for leakage and replaced occasionally.

The $\dsThreeDOne$ and $\ddThreePOne$ manifolds are also connected by a strong electric quadrupole transition at $\lambda \approx 759\,\nm$, which would allow a ``background-free'' measurement scheme, i.e. one which avoids applying lasers at the detected wavelength~\cite{Roman2020}, by driving transitions $\spThreePZero$ $\leftarrow$ $\ddThreePOne$ $\leftarrow$ $\dsThreeDOne$ and measuring fluorescence at $\approx 442\nm$. 

\subsubsection{$\dpThreeFFour$ measurement}

The $\dpThreeFFour$ level has a much larger linewidth of $\approx 20\,\MHz$, but with several decay channels, to $\dsThreeDThree$,  $\ddThreeFThree$, $\ddThreeFFour$, and $\ddOneGFour$ (calculated lifetimes $853.5\,\s$, $12.3\,\s$, $11.3\,\s$, $1.4\,\s$, and branchings $88.6\%$, $0.73\%$, $10.6\%$, $0.0004\%$, respectively). Fig.~\ref{fig:measurement_combo}b illustrates these transitions. Measurement using $\dpThreeFFour$ would allow for the suppression of crosstalk for qubits stored in the $\dsThreeDOne$ clock states or in the ground-state manifold. Several quadrupole transitions among these levels could potentially allow for background-free measurement schemes. 

Assuming a population of 0.1 in the shortest-lifetime manifold requiring repumping ($\ddOneGFour$) would allow for a measurement time of $1.4\,\ms$, similar to the $\spThreePZero$ measurement scheme, but with an $8\times$ larger bare cycling transition rate.

Note we have not calculated the out-of-cycle branching rates due to non-dipole decays or hyperfine quenching for this method. However, direct non-dipole decays from $\dpThreeFFour$ into the qubit manifolds $\dsThreeDOne$ or $\ssOneSZero$ should be insignificant due to the large difference in $J$. 
For hyperfine mixing, nearby excited levels either have a large difference in $J$ from $\dpThreeFFour$, which would require higher-order hyperfine interactions to mix and should therefore be suppressed; or again have a large difference in $J$ from the qubit manifolds. A large energy separation from $4d5p\ {}^3\!F_4$ ($\geq 24$ THz) further
 suppresses the hyperfine mixing amplitude itself. Therefore, we expect that any additional decays will be small and not into the qubit storage manifolds.

\subsection{Single-qubit gates} \label{sec:SQ}

\subsubsection{Laser-based gates}

\begin{figure}
 \includegraphics[width=1.0\columnwidth]{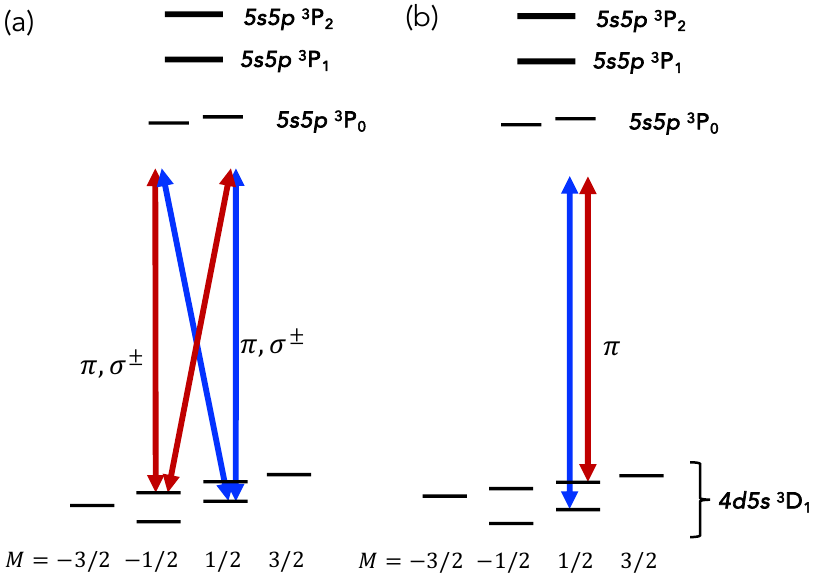}
 \includegraphics[width=0.9\columnwidth]{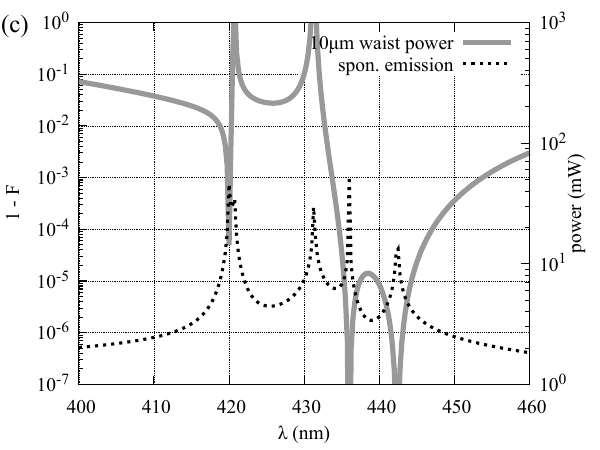}
 \caption{\label{fig:laser_SQ} Laser-based Raman single-qubit gates in the $\dsThreeDOne$ manifold. (a) Transitions for a gate on the states targeted by the circularly polarized shelving technique of Sec.~\ref{sec:shelving}, at $400\,\Gauss$. A common linear polarization providing $\sigma^+$, $\sigma^-$, and $\pi$ couplings (e.g., angled $45^\circ$ with respect to the quantization axis) is used for both tones. (b) Transitions for a gate on the hyperfine clock qubit at $B\approx 168\,\Gauss$, using linear polarization parallel to the quantization axis. (c) Estimated spontaneous emission error and required laser power vs. wavelength for the $\pi$-polarized clock qubit single-qubit gate. Results for the inner states (of panel (a)) look very similar.}
\end{figure}

One method to perform single-qubit gates is to drive Raman transitions between qubit states in the $\dsThreeDOne$ manifold using the $\dsThreeDOne \leftrightarrow \spThreePZero$ transition. This could be done either on the $\dsThreeDOne,\,M=\pm 1/2$ states targeted by the circularly polarized shelving scheme of Sec.~\ref{sec:shelving}, or on the clock qubit states. Conveniently, selection rules allow a common linear polarization to be used for the two tones, unlike the typical alkaline earth $S_{1/2}, M=0$ clock qubits. Moreover, the state splitting is at most a few hundreds of $\MHz$.  Fig.~\ref{fig:laser_SQ} illustrates these options.

At larger bias fields, the $\dsThreeDOne,\,M=\pm 1/2$ states acquire greater nuclear spin purity.
Although this improves the coherent shelving procedure of Fig.~\ref{fig:circular_shelving_combo}, it can lead to increased intensity requirements for single-qubit gates between the shelved states. However, the power required for a high-fidelity gate is still generally quite manageable. For example, at $B=400\,\Gauss$, we estimate a $10\,\mu\s$ $\pi$ rotation (with $1\,\us$ $\sin^2$ turn-on and turn-off) between the $M=\pm 1/2$ shelved states at $\lambda=445\nm$ can be accomplished with $\approx 18\mW$ of power in a Gaussian beam with $10\mu\meter$ beam waist radius, with spontaneous emission error of $\approx 3\times 10^{-6}$. This is assuming two co-propagating tones in a single beam with linear polarization angled $45^\circ$ from the quantization axis. For the $\dsThreeDOne$ clock qubit states at $B \approx 168\Gauss$, with $\pi$-polarized light, the same gate achieves $\approx 2\times 10^{-6}$ spontaneous emission error with a power of $\approx 10\mW$. Fig.~\ref{fig:laser_SQ}c shows the wavelength dependence of this gate for the clock qubit states (the wavelength dependence for the $M=\pm1/2$ states at $B=400\,\Gauss$ looks very similar, with slightly larger laser power requirements and errors).

\subsubsection{Magnetic gates}\label{sec:magnetic_gates}

The nuclear spin qubit $\ssOneSZero$ is amenable to magnetic gates, however, its small magnetic moment would require large magnetic fields or long pulse times. Instead, single-qubit magnetic gates could be implemented in a metastable manifold, which additionally protects spectator nuclear spin qubits from magnetic field crosstalk.
In the $\dsThreeDOne$ manifold, magnetic single-qubit gates could be implemented with fields oscillating at a few hundreds of $\MHz$ or less.
For the clock qubit states, the magnetic dipole transition strength is $\approx 0.25\mu_B$, which is somewhat smaller than the $\approx 1.4\mu_B$ of a (field-sensitive) spin-1/2 Zeeman qubit (e.g., ${}^{40}\mathrm{Ca}^+$).
Nonetheless, a $\pi$-rotation between these states could be accomplished in $10\us$ with a magnetic field of amplitude $\approx 0.15\Gauss$ oscillating at $\approx 330\MHz$.

\begin{figure}
\includegraphics[width=.75\columnwidth]{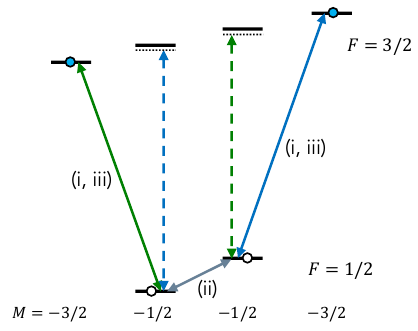}
\caption{\label{fig:3D1_magnetic}Transitions for performing magnetic single-qubit gates on a qubit temporarily shelved in the $\dsThreeDOne, M=\pm3/2$ states. Dashed arrows indicate near-degenerate transitions which can be avoided with sufficiently large magnetic field and/or sufficiently good polarization control. Level structure is shown at $B=50\Gauss$ (off-resonant detunings are exaggerated for visibility).}
\end{figure}

For qubits temporarily shelved into the $\dsThreeDOne$, $M=\pm 3/2$ states at intermediate bias fields, Fig.~\ref{fig:3D1_magnetic} shows how single-qubit magnetic gates could be accomplished with multiple pulses by (i) simultaneously mapping the $F=3/2, M=\pm3/2$ states to the $F=1/2$, $M=\pm 1/2$ states, then (ii) applying a gate pulse, then (iii) un-mapping, with similar field strength requirements to the clock qubit. The pulses (i), (ii), and (iii) are indicated in the diagram. Near-degeneracies between $F=3/2,M=3/2 \leftrightarrow F=1,M=1/2$ and $F=3/2,M=-1/2 \leftrightarrow F=1/2,M=-1/2$, as well as between $F=3/2,M=-3/2 \leftrightarrow F=1,M=-1/2$ and $F=3/2,M=1/2 \leftrightarrow F=1/2,M=1/2$, can be avoided with careful polarization control of the magnetic field and/or a sufficiently large bias field. E.g., at $B=50~\Gauss$, the degeneracies are broken by $\gtrsim 680~\kHz$, while at $B=100~\Gauss$ they are broken by $\gtrsim 2.4~\MHz$.

\subsection{Two-qubit gates}
The level structure of $\Y$ allows for many potential two-qubit gate implementations. 
Examples include light-shift or Mølmer-Sørensen gates on the dipole transition between $\dsThreeDOne$ and $\spThreePZero$, light-shift gates on one of several electric quadrupole transitions~\cite{Baldwin2021,Sawyer2021,Clark2021}, microwave-driven Mølmer-Sørensen gates on the clock qubit, or a radiowave-driven $ZZ$ gate on a field-sensitive pair of states~\cite{Ospelkaus2008}. In this section we discuss a few possibilities, focusing on implementations within the $\dsThreeDOne$ manifold.

\subsubsection{Laser-based gates}
As one example of a laser-based two-qubit gate, we estimate spontaneous emission errors and required intensity for a light-shift gate on the dipole transition to $\spThreePZero$ at $B=400\,\Gauss$, with one qubit state in $\ssOneSZero,\,M=-1/2$ and the other shelved to the lower-energy $\dsThreeDOne,\,M=+1/2$ state. This is similar to the gate described in Refs.~\cite{Sawyer2021, Clark2021}.

\begin{figure}
\smallskip
\includegraphics[width=.7\columnwidth]{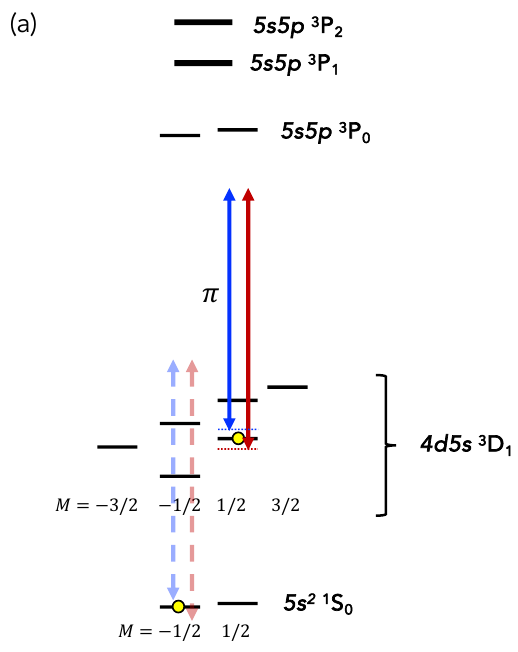}
\includegraphics[width=1.0\columnwidth]{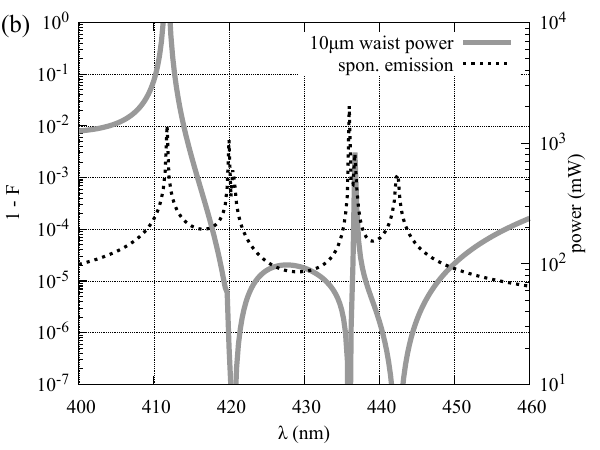}
\caption{\label{fig:piLS_diagram_combo} (a) Transition diagram for a two-qubit light-shift gate using $\pi$-polarized beams on qubits in the  $\ssOneSZero, M=-1/2$ and $\dsThreeDOne, M=+1/2$ states. (b) Estimated errors and beam intensity for a two-qubit light-shift gate using $\pi$-polarized beams on qubits in the  $\ssOneSZero, M=-1/2$ and $\dsThreeDOne, M=+1/2$ states.}
\end{figure}

Fig.~\ref{fig:piLS_diagram_combo}a shows the transition diagram for the gate. 
The $\dsThreeDOne,\,M=+1/2$ state couples strongly to the $\spThreePZero$ manifold, while coupling between $\ssOneSZero$ and $\spThreePZero$ is suppressed due to selection rules. 

Fig.~\ref{fig:piLS_diagram_combo}b shows calculated spontaneous emission errors and power requirements for this gate as a function of wavelength $\lambda$. We assume counter-propagating beams are directed axially along a $\Y\textrm{-}\Sr\textrm{-}\Sr\textrm{-}\Y$ crystal with axial center-of-mass and out-of-phase mode frequencies of $1.0\,\MHz$ and $1.73\,\MHz$, respectively. 
For wavelength $\lambda=450\,\nm$, this gives a Lamb-Dicke parameter of $\eta = 0.108$ for coupling to the out-of-phase mode. 
We consider a two-loop gate with a total time of each loop of $25\,\mu \s$, including a smooth $\sin^2$ turn-on and turn-off each of $4\,\us$, and assume a Gaussian beam waist radius of $10\,\um$ for power estimates. 
A spin-echo pulse would likely be desired between the two loops (as in, e.g., Refs.~\cite{Baldwin2021,Sawyer2021}), but this is omitted from the simulation, as it should contribute negligibly to the spontaneous emission errors. 
The calculation of spontaneous emission errors is described in the Appendix (note that we do not include the effects of motional dephasing due to recoil). 
For these parameters, we find that an error of $\approx 10^{-5}$ can be achieved with $\approx 100\mathrm{mW}$ of laser power at $\lambda\approx 450\nm$.

\subsubsection{Magnetic gradient gates}\label{sec:magnetic_gradient_gates}
Ref.~\cite{Ospelkaus2008} proposed a magnetic-gradient multi-qubit gate which is executed by temporarily mapping qubits into magnetic-field-sensitive states and applying a magnetic gradient oscillating at a frequency close to that of a motional mode. This generates an interaction Hamiltonian of the form $\hat{H} \approx (\Omega/2)\sum_q \sigma_z^{(q)} (a_g^\dagger e^{-i \delta t} + a_g e^{i \delta t})$, where $q=1,\ldots,N$ indexes the qubits, $a_g$ and $a^\dagger_g$ are the gate motional mode lowering and raising operators, and $\delta$ is the detuning from resonance with the mode. The phase space dynamics are similar to the Mølmer-Sørensen gate.

The $\dsThreeDOne$, $M=\pm 3/2$ states provide a convenient choice for a magnetic gradient gate, with relative field sensitivity $d\omega_q/dB \approx 2\pi \cdot 1400\kHz/\Gauss$ and a robust shelving scheme from the nuclear spin qubit described in Sec.~\ref{sec:shelving}. For a bias field along $z$ and a gate coupling to a mode in the $y$ direction, and applied gradient $(\partial B_z/\partial y) \cos(\omega t)$, the strength of the coupling is given by 
\begin{equation}
    \Omega = \frac{1}{2} \frac{\partial\omega_q}{\partial B} \frac{\partial B_z}{\partial y} \frac{y_0}{\sqrt{2}}
\end{equation} 
where $y_0$ is the oscillator length for the qubit ion motion in the normal mode of the gate ($\hat{y} = y_0 (a_g + a^\dagger_g)/\sqrt{2}$).
For these states, assuming $y_0 = 6\nm$, a single-loop maximally entangling gate (for which $\Omega = \pi/T$) can be executed in $50\,\us$ with a gradient of magnitude $\approx 340\,\Tesla/\meter$. A spin-echo pulse between these states could be implemented with magnetic fields using the pulse sequence described in Sec.~\ref{sec:magnetic_gates} (which would require a field component perpendicular to the bias field).

Alternatively, a Mølmer-Sørensen gate can be implemented on a pair of clock states, at the cost of requiring a higher-frequency gradient oscillation (tuned close to a first-order motional sideband transition between the states). For alkaline earth elements, clock frequencies in the ground-state manifold are generally on the order of several $\GHz$. The clock frequency in $\dsThreeDOne$ is much smaller, $\approx 330\MHz$. In this case, for the same applied bias field direction and gate mode as before, the strength of the coupling is
\begin{equation}
    \Omega = \mu_{\uparrow,\downarrow} \frac{\partial B_z}{\partial y} \frac{y_0}{\sqrt{2}}
\end{equation}
where $\mu_{\uparrow,\downarrow}$ is the M1 transition matrix element between the clock states (here the gate couples to $B_z$ because the clock states have the same magnetic quantum number). 
For these states, we calculate $\mu_{\uparrow,\downarrow} \approx 0.25 \mu_B$, and a $50\us$ single-loop Mølmer-Sørensen gate would require a gradient of magnitude $\approx 680 \,\Tesla/\meter$.

\section{Conclusion}
\label{sec:conclusion}

Despite the leading performance of trapped ion quantum computers, relatively few atomic elements have been explored as qubit hosts, with the overwhelming majority of work performed using ions with a single valence electron in an $s$ orbital.
As a result, much of the periodic table remains essentially unexplored and there is potential for significant optimization of qubit performance.
Here, we have taken a step towards this goal by exploring $\Y$ both experimentally and theoretically.
Employing cryogenically-cooled $\Y$ produced via laser ablation, we recorded high-resolution spectroscopy with laser-induced fluorescence to measure branching ratios and the hyperfine coefficients of several manifolds expected to be used in quantum operations.
Further, we performed atomic structure calculations of lifetimes, hyperfine coefficients, transition rates, and matrix elements ($E1$, $E2$, and $M1$) for all manifolds up to and including $\dpThreeDThree$. 

We find that the multivalent structure of $\Y$ leads to several new resources for quantum information processing, including a nuclear qubit, several classes of metastable manifolds based on different electronic configurations, and low-crosstalk state preparation, measurement, and gate operations. 
The $I=1/2$ nuclear spin of $\Y$ provides a nuclear qubit in the ground $\ssOneSZero$ configuration and is nearly ideal for quantum information storage.
This qubit has an attractive combination of small sensitivity to slow bias field fluctuations and to higher-frequency oscillating magnetic fields -- e.g., at frequencies used for ion trapping or magnetic gates. 
Above this ground-state manifold, there are many metastable manifolds arising from the $4d5s$ and $4d^2$ configurations, which are calculated to have long lifetimes; the $\dsThreeDOne$ has an estimated lifetime of $\sim 10^{10}\,\s$, and the next two lowest-energy manifolds have estimated lifetimes of $\sim 10^3\,\s$. 
These offer finite-field hyperfine clock qubits at much smaller magnetic fields and qubit frequencies than are typical in alkaline earth ions.
Further, calculations show that the nuclear spin qubit can be coherently and robustly shuttled to these metastable manifolds for implementing both laser-based and laser-free gates.
For example, the stretched $\dsThreeDOne, M=\pm 3/2$ states provide an attractive manifold for implementing a magnetic-gradient $ZZ$ gate.
Finally, the $\dsThreeDOne$ level also has a nearly closed cycling transition to $\spThreePZero$, allowing for laser cooling and SPAM that are also highly-decoupled from the nuclear storage qubit. An additional cycling transition between $\dsThreeDThree$ and $\dpThreeFFour$ could allow these operations to be decoupled from a hyperfine qubit stored in $\dsThreeDOne$.

$\Y$ therefore appears especially amenable to \textit{omg}-style operation with quantum information stored in the nuclear spin degree of freedom while gating and entropic operations are performed in metastable manifolds. It also supports alternative strategies in which information is stored in long-lived metastable clock qubits with much smaller transition frequencies than those found in the ground-state manifolds of alkaline earth ions.

Practical, large-scale trapped-ion quantum computers must not only achieve high-fidelity operations on targeted qubits but also satisfy constraints on crosstalk and resource usage (such as laser power).
The multivalent structure of $\Y$ provides new opportunities to meet these demands. 
In particular, the separation of storage and operation manifolds should minimize qubit crosstalk in large-scale machines during entropic operations. 
Further, the same separation appears to solve one of the chief challenges of laser-free gates: crosstalk during gates.
Because the nuclear spin qubit is largely insensitive to magnetic fields, qubits can be mapped into metastable states, such as the magnetically-sensitive $\dsThreeDOne$ state, for gating.
Thus, even if the applied magnetic field gradients extend beyond the gating region, quantum information stored in the nuclear qubit remains intact. 
Together, these features suggest $\Y$ allows for the construction of a quantum computer with uniquely small memory errors and crosstalk, while offering resource-efficient schemes for quantum logic gates.

\section*{Acknowledgments}
The authors thank Stephen Erickson, John P. Gaebler, Sergey Porsev, and Matthew Swallows for helpful discussions. 
This work has been supported in part by the US Office of Naval Research 
US Office of Naval Research Grant N000142512105, the National Science Foundation (PHY-2309254, PHY-2110421, and OMA-2016245) and the Army Research Office (W911NF-20-1-0037).
ERH acknowledges support from the NSF NQVL (OSI-2435382).

\bibliography{Reference}

\clearpage

\appendix
\section{Reduced matrix elements}\label{appendix:electronic_structure}

\onecolumngrid
\begingroup
\setlength{\LTcapwidth}{7.0in} 
\setlength{\extrarowheight}{2.8pt}
\setlength\tabcolsep{8.8pt}

\begin{longtable*}{lccccccccc}   

\label{trans_tab} \vspace{-0pt}    \\
    \hline\hline
Upper & Term &Level& Lower & Term & Level& $\lambda$& & <Up||E,M||Low>& Tr. rate \\
\hline
\endfirsthead
\multicolumn{10}{c}%
{{\tablename\ \thetable{} -- continued from the previous page}} \vspace{+2pt}\\
\hline\hline
Upper & Term &Level& Lower & Term & Level& $\lambda$& & <Up||E,M||Low>& Tr. rate \\
\hline
\endhead
\hline
\endfoot
\hline \hline
\endlastfoot
4d5s	        &	a $^3$D$_1$	&	840.2 &	5s$^2$&	a $^1$S$_0$	&	0.0  & 11902.0 &	M1	&	6.5(4)$\times 10^{-5}$ & 2.22(25)$\times 10^{-11}$ \\
                &	a $^3$D$_2$	&  1045.1 &	5s$^2$&	a $^1$S$_0$	&	0.0  & 9568.7  &    E2	&	-1.19(6) & 3.9(4)$\times 10^{-8}$ \\
                &	            &         &	4d5s  &	a $^3$D$_1$	& 840.2  & 48809.5 &	M1	&	-2.1035(24)& 2.053(5)$\times 10^{-4}$ \\     &	            &         &	      &	            &        &         &	E2	&	-4.15(7)& 1.39(5)$\times 10^{-10}$ \\     
4d5s	        &	a $^3$D$_3$	&  1449.8 &	4d5s  &	a $^3$D$_1$	& 840.2  & 16405.4 &	E2	&	1.410(21) & 2.68(8)$\times 10^{-9}$ \\
                &	            &         &	4d5s  &	a $^3$D$_2$	&1045.1  & 24711.1 &	M1	&	2.1421(25)& 1.1716(27)$\times 10^{-3}$ \\    &	            &         &	      &	            &        &         &	E2	&	4.46(8)& 3.46(13)$\times 10^{-9}$ \\
4d5s	        &	a $^1$D$_2$	&  3296.2 &	5s$^2$&	a $^1$S$_0$	&	0.0  & 3033.8  &	E2	&	-9.49(17) & 7.85(28)$\times 10^{-4}$ \\
                &	            &         &	4d5s  &	a $^3$D$_1$	& 840.2  &	4071.7 &    M1	&	0.272(18)& 5.9(8)$\times 10^{-3}$ \\
                &	            &         &	      &	            &        &         &	E2	&	0.315(6)& 1.98(8)$\times 10^{-7}$ \\ 
                &	            &         &	4d5s  &	a $^3$D$_2$	&1045.1  &	4442.3 &    M1	&	-0.116(8)& 8.3(1.1)$\times 10^{-4}$ \\
                &	            &         &	      &	            &        &         &	E2	&	1.37(13)& 2.4(5)$\times 10^{-6}$ \\ 
                &	            &         &	4d5s  &	a $^3$D$_3$	&1449.8  &	5415.9 &    M1	&	-0.277(19)& 2.6(3)$\times 10^{-3}$ \\
                &	            &         &	      &	            &        &         &	E2	&	-0.58(4)& 1.60(23)$\times 10^{-7}$ \\      
4d$^2$	        &	a $^3$F$_2$	&  8003.1 &	5s$^2$&	a $^1$S$_0$	&	0.0  & 1249.5  &	E2	&	-0.407(25) & 1.22(15)$\times 10^{-4}$ \\
                &	            &         &	4d5s  &	a $^3$D$_1$	& 840.2  &	1396.1 &    M1	&	0.0052(28)& $\leq1.1\times 10^{-4}$ \\
                &	            &         &	      &	            &        &         &	E2	&	9.96(11)  & 4.19(9)$\times 10^{-2}$ \\ 
                &	            &         &	4d5s  &	a $^3$D$_2$	&1045.1  &  1437.2 &    M1  &   0.00123(16)& 2.7(7)$\times 10^{-6}$ \\
                &	            &         &	      &	            &        &  1437.2 &    E2  &   -8.34(12)& 2.54(7)$\times 10^{-2}$ \\
                &	            &         &	4d5s  &	a $^3$D$_3$	&1449.8  &	1525.9 &    M1	&	-0.0016(29)& $\leq1.8\times 10^{-5}$ \\
                &	            &         &	      &	            &        &         &	E2	&	2.6(4)& 1.80(10)$\times 10^{-3}$ \\  
                &	            &         &	4d5s  &	a $^1$D$_2$	&3296.2  &	2124.5 &    M1	&	0.055(7)& 1.7(4)$\times 10^{-3}$ \\
                &	            &         &	      &	            &        &         &	E2	&	0.78(11)& 3.2(9)$\times 10^{-5}$ \\  
4d$^2$	        &	a $^3$F$_3$	&  8328.0 &	4d5s  &	a $^3$D$_1$	& 840.2  & 1335.5  &	E2	&	-7.09(8) & 1.90(4)$\times 10^{-2}$ \\
                &	            &         &	4d5s  &	a $^3$D$_2$	&1045.1  &  1373.1 &    M1  &   -0.0080(15)& 9(4)$\times 10^{-5}$ \\
                &	            &         &	      &	            &        &         &    E2  &   -11.10(15)& 4.04(11)$\times 10^{-2}$ \\
                &	            &         &	4d5s  &	a $^3$D$_3$	&1449.8  &1453.9 & M1 &3.330(20)$\times10^{-3}$& 1.390(17)$\times 10^{-5}$\\
                &	            &         &	      &	            &        &         &	E2	&	8.56(10)& 1.81(4)$\times 10^{-2}$ \\ 
                &	            &         &	4d5s  &	a $^1$D$_2$	&3296.2  &	1987.3 &    M1	&	-0.078(9)& 3.0(7)$\times 10^{-3}$ \\
                &	            &         &	      &	            &        &         &	E2	&	1.23(7)& 7.8(9)$\times 10^{-5}$ \\ 
                &	            &         &	4d$^2$&	a $^3$F$_2$ &8003.1  & 30777.5 &    M1	&	-2.578(24)& 8.785(16)$\times 10^{-4}$ \\
                &	            &         &	      &	            &        &         &	E2	&	-1.855(12)& 1.993(25)$\times 10^{-10}$ \\
4d$^2$	        &	a $^3$F$_4$	&  8743.3 &	4d5s  &	a $^3$D$_2$	& 1045.1  & 1299.0 &	E2	&	-7.34(8) & 1.81(4)$\times 10^{-2}$ \\
                &	            &         &	4d5s  &	a $^3$D$_3$	&1449.8  &1371.1 & M1 &3.680(20)$\times10^{-3}$& 1.575(17)$\times 10^{-5}$\\
                &	            &         &	      &	            &        &         &	E2	&	-16.33(19)& 6.85(16)$\times 10^{-2}$ \\ 
                &	            &         &	4d5s  &	a $^1$D$_2$	&3296.2  &	1835.8 &    E2	&	0.64(5)& 2.4(4)$\times 10^{-5}$ \\
                &	            &         &	4d$^2$&	a $^3$F$_2$ &8003.1  & 13509.9 &    E2	&	-0.360(4)& 3.59(9)$\times 10^{-10}$ \\
                &	            &         &	4d$^2$&	a $^3$F$_3$ &8328.0  & 24080.0 &    M1	&	-2.597(3)& 1.447(3)$\times 10^{-3}$ \\
                &	            &         &	      &	            &        &         &	E2	&	-1.873(24)& 5.39(14)$\times 10^{-10}$ \\    
4d$^2$	        &	a $^3$P$_0$	& 13883.4 &	5s$^2$&	a $^3$D$_1$	& 840.2  & 766.7   &	M1&1.00(10)$\times10^{-4}$& 6.0(1.7)$\times10^{-7}$\\
                &	            &         &	4d5s  &	a $^3$D$_2$	&1045.1  &  778.9  &    E2  &   -6.00(5)& 1.404(24) \\
                &	            &         &	4d5s  &	a $^1$D$_2$	&3296.2  &	944.5  &    E2	&	0.76(6)& 8.6(1.3)$\times 10^{-3}$ \\
                &	            &         &	4d$^2$&	a $^3$F$_2$ &8003.1  & 1700.6  &    E2	&	-4.88(6)& 1.88(5)$\times 10^{-2}$ \\  
4d$^2$	        &   a $^3$P$_1$	& 14018.3 &5s$^2$ &	a $^1$S$_0$	&	0.0  &	713.4  &    M1  &  0.0037(9)& 3.4(17)$\times10^{-4}$\\
                &	            &         &	4d5s  &	a $^3$D$_1$	& 840.2  &	758.8  & M1&2.40(20)$\times 10^{-4}$& 1.19(20)$\times 10^{-6}$\\
                &	            &         &	      &	            &        &         &    E2  & 7.08(7)	&7.43(14)$\times 10^{-1}$ \\
                &	            &         &	4d5s  &	a $^3$D$_2$	&1045.1  &	770.8  &    M1  &    0.0035(19)	&$\leq5\times 10^{-4}$ \\
                &	            &         &	      &	            &        &         &    E2  & 2.994(26)	&1.230(21)$\times 10^{-1}$ \\
                &	            &         &	4d5s  &	a $^3$D$_3$	&1449.8  &	795.6  &    E2  &    -7.11(7)	&5.91(11)$\times 10^{-1}$ \\
                &	            &         &	4d5s  &	a $^1$D$_2$	&3296.2  &	932.7  &    M1  &  0.01747(21)	&3.38(8)$\times 10^{-3}$  \\
                &	            &         &	      &	            &        &	       &    E2  &  -0.300(5)	&4.76(15)$\times 10^{-4}$	\\
                &	            &         &	4d$^2$&	a $^3$F$_2$ &8003.1  &	1662.5 &    M1  &  0.00185(10)&6.7(7)$\times 10^{-6}$ \\
                &	            &         &	      &	            &        &	       &    E2  &  4.90(5)	&7.07(15)$\times 10^{-3}$ \\
                &	            &         &	4d$^2$&	a $^3$F$_3$ &8328.0  &	1757.4 &    E2  &  6.89(8)	&1.057(25)$\times 10^{-2}$ \\
                &	            &         &	4d$^2$&	a $^3$P$_0$	& 13883.4&	74135.6&    M1  &  -1.413(4)	&4.403(25)$\times 10^{-5}$ \\            
4d$^2$	        &	a $^3$P$_2$	& 14098.1 &	5s$^2$&	a $^1$S$_0$	&	0.0 & 	709.3&    E2  &1.16(3)    &1.68(10)$\times 10^{-2}$\\
                &               &         &	4d5s&	a $^3$D$_1$   & 840.2  &754.3	 &    M1  &-0.0073(24)&7(4)$\times 10^{-4}$  \\
                &               &         &	      &	            &        &       &    E2  &3.99(6)	  &1.46(4)$\times 10^{-1}$   \\
                &               &         &	4d5s&	a $^3$D$_2$   &1045.1  &766.1  &    M1  &0.0091(20) &1.0(4)$\times 10^{-3}$  \\
                &               &         &	      &	            &        &       &    E2  &7.92(8)	  &5.32(11)$\times 10^{-1}$  \\
                &               &         &	4d5s&	a $^3$D$_3$   &1449.8  &790.6  &    M1  &0.0074(25) &6(4)$\times 10^{-4}$  \\
                &               &         &	      &	            &        &       &    E2  &9.43(13)	  &6.44(18)$\times 10^{-1}$  \\
                &               &         &	4d5s&	a $^1$D$_2$   &3296.2  &925.8  &    M1  &0.0287(5)  &5.6(2)$\times 10^{-3}$  \\
                &               &         &	      &	            &        &       &    E2  &3.744(12)  &4.62(3)$\times 10^{-2}$  \\
                &               &         &	4d$^2$&	a $^3$F$_2$   &8003.1  &1640.7 &    M1  &0.0339(4)  &1.40(3)$\times 10^{-3}$ \\
                &               &         &	      &	            &        &       &    E2  &1.57(3)    &4.63(20)$\times 10^{-4}$\\
                &               &         &	4d$^2$&	a $^3$F$_3$   &8328.0  &1733.1 &    M1  &-0.0438(8) &1.99(7)$\times 10^{-3}$  \\
                &               &         &	      &	            &        &       &    E2  &4.63(7)    &3.08(9)$\times 10^{-3}$  \\
                &               &         &	4d$^2$&	a $^3$F$_4$   &8743.3  &1867.5 &    E2  &9.10(15)   &8.17(28)$\times 10^{-3}$ \\
                &               &         &	4d$^2$&	a $^3$P$_0$   &13883.4&46577.3 &    E2  &4.62(7)    &2.18(6)$\times 10^{-10}$ \\
                &               &         &	4d$^2$&	a $^3$P$_1$   &14018.3&125299.2&    M1  &1.472(8)   &5.94(7)$\times 10^{-6}$ \\
                &               &         &	      &	            &        &       &    E2  &-6.88(10)  &3.43(10)$\times 10^{-12}$\\
4d$^2$	        &	b $^1$D$_2$	& 14832.9 &	5s$^2$&	a $^1$S$_0$	&	0.0  & 674.2 &    E2  &-2.6825(28)&1.1572(24)$\times 10^{-1}$ \\
                &               &         &	4d5s&	a $^3$D$_1$   & 840.2  &	714.7&    M1  &0.0236(19) &	8.2(13)$\times 10^{-3}$ \\
                &               &         &	    &               &        &       &    E2  & 1.12(5)   &	1.50(13)$\times 10^{-2}$ \\
                &               &         &	4d5s&	a $^3$D$_2$   &1045.1  &725.3  &    M1  &-0.00786(26)&	8.7(6)$\times 10^{-4}$\\
                &               &         &	    &               &        &       &    E2  &	1.6566(21)&	3.063(8)$\times 10^{-2}$\\
                &               &         &	4d5s&	a $^3$D$_3$   &1449.8  &747.2  &    M1  &	-0.0243(20)&	7.6(12)$\times 10^{-3}$\\
                &               &         &	    &               &        &       &    E2  &	3.55(10)&	1.21(7)$\times 10^{-1}$\\
                &               &         &	4d5s&	a $^1$D$_2$   &3296.2  &866.8  &    M1  &	0.01009(10)&8.43(17)$\times 10^{-4}$\\
                &               &         &	    &               &        &       &    E2  &	-12.3(3)&7.0(3)$\times 10^{-1}$\\
                &               &         &	4d$^2$&	a $^3$F$_2$   &8003.1  &1464.2 &    M1  &	-0.0706(13)&	8.6(3)$\times 10^{-3}$\\
                &               &         &	    &	            &        &       &    E2  &	1.21(6)&	4.9(5)$\times 10^{-4}$\\
                &               &         &	4d$^2$&	a $^3$F$_3$   &8328.0  &1537.3 &    M1  &	0.1015(19)&	1.53(6)$\times 10^{-2}$\\
                &               &         &	    &	            &        &       &    E2  &	1.75(5)&	8.0(5)$\times 10^{-4}$\\
                &               &         &	4d$^2$&	a $^3$F$_4$   &8743.3  &1642.2 &    E2  &	3.21(7)&	1.93(9)$\times 10^{-3}$\\
                &               &         &	4d$^2$&	a $^3$P$_0$   &13883.4&10532.0 &    E2  &	1.57(3)&	4.24(16)$\times 10^{-8}$\\
                &               &         &	4d$^2$&	a $^3$P$_1$   &14018.3&12276.0 &    M1  &	0.576(20)&	9.7(7)$\times 10^{-4}$\\
                &               &         &	    &	            &        &       &    E2  &	-2.48(7)&	4.92(28)$\times 10^{-8}$\\
                &               &         &	4d$^2$&	a $^3$P$_2$   &14098.1&13609.4 &    M1  &	0.930(28)&	1.85(11)$\times 10^{-3}$\\
                &               &         &	    &	            &        &       &    E2  &	5.57(19)&	1.49(10)$\times 10^{-7}$\\
4d$^2$	        &	a $^1$G$_4$	& 15682.9 &	4d5s&	a $^3$D$_2$   &1045.1  & 683.2 &	  E2  &	-1.83(8) & 2.80(23)$\times 10^{-2}$ \\
                &               &         &	4d5s&	a $^3$D$_3$   &1449.8  &702.6  &    M1  &-1.9$\times 10^{-4}$	&3.12$\times 10^{-7}$\\
                &               &         &	    &	            &        &       &    E2  & 0.5728(5) & 2.385(4)$\times 10^{-6}$\\
                &               &         &	4d5s&	a $^1$D$_2$   &3296.2  &807.3  &    E2  &-13.7(3)	& 6.8(3)$\times 10^{-1}$\\
                &               &         &	4d$^2$&	a $^3$F$_2$   &8003.1  &1302.1 &    E2  &1.06(3) &	3.76(21)$\times 10^{-4}$\\
                &               &         &	4d$^2$&	a $^3$F$_3$   &8328.0  &1359.6 &    M1  &0.0785(10) &	7.34(19)$\times 10^{-3}$\\
                &               &         &	    &	            &        &       &    E2  &0.1645(6)&7.24(5)$\times 10^{-6}$\\
                &               &         &	4d$^2$&	a $^3$F$_4$   &8743.3  &1441.0 &    M1  &-0.1012(13) &	1.025(26)$\times 10^{-2}$\\
                &               &         &	    &	            &        &       &    E2  &-0.43373(5)&3.7674(12)$\times 10^{-5}$\\
                &               &         &	4d$^2$&	a $^3$P$_2$   &14098.1 &6309.8 &    M1  &5.17(17)&	3.32(22)$\times 10^{-6}$\\
                &               &         &	4d$^2$&	b $^1$D$_2$ &14832.9  &11764.1&   E2  &-13.66(10) &	1.030(15)$\times 10^{-6}$\\

5s5p            &	z $^3$P$^o_0$&23445.1 &	4d5s&	a $^3$D$_1$ & 840.2   & 442.4&	E1	&	0.87(4) & 1.76(15)$\times 10^{7}$ \\
                &	             &        &	4d$^2$&	a $^3$P$_1$ & 14018.3& 1060.8&	E1	&	0.746(9) & 9.45(23)$\times 10^{5}$ \\
5s5p            &	z $^3$P$^o_1$&23776.2 &	5s$^2$&	$^1$S$_0$	&	0.0 &  420.6 &	E1	&	0.48(4) & 2.10(35)$\times 10^{6}$ \\
                &	             &        &	4d5s&	a $^3$D$_1$ & 840.2 &  436.0 &	E1	&	0.799(18) & 5.20(23)$\times 10^{6}$ \\
                &	             &        &	4d5s&	a $^3$D$_2$ &1045.1  & 439.9 &	E1	&	1.17(8) & 1.08(15)$\times 10^{7}$ \\ 
                &	             &        &	4d5s&	a $^1$D$_2$ &3296.2  & 488.3 &	E1	&	0.149(4) & 1.30(7)$\times 10^{5}$ \\ 
                &	             &        &	4d$^2$&	a $^3$F$_2$ &8003.1  & 634.0 &	E1	&	0.112(23)& 3.4(1.4)$\times 10^{4}$ \\
                &	            &         &	4d$^2$&	a $^3$P$_0$	& 13883.4& 1010.8&  E1  &   0.785(13)& 4.03(13)$\times 10^{5}$ \\
                &               &         &	4d$^2$&	a $^3$P$_1$ &14018.3 & 1024.8&  E1  &	0.639(7)&	2.56(5)$\times 10^{5}$\\
                &               &         &	4d$^2$&	a $^3$P$_2$ &14098.1& 1033.3 &  E1  &	0.925(22)&	5.24(25)$\times 10^{5}$\\
                &               &         &	4d$^2$&	b $^1$D$_2$ &14832.9  &1118.1&  E1  &  0.055(18) &	1.4(1.0)$\times 10^{3}$\\
                
5s5p            &	z $^3$P$^o_2$&24647.1 &	4d5s&	a $^3$D$_1$ & 840.2   & 420.0&	E1	&	0.30(10) & 5(3)$\times 10^{5}$ \\
                &	             &        &	4d5s&	a $^3$D$_2$ &1045.1  & 423.7 &	E1	&	0.60(17)(8) & 1.9(1.1)$\times 10^{6}$ \\
                &               &         &	4d5s&	a $^3$D$_3$ &1449.8  & 431.1 &  E1  & 1.53(13)     &1.19(20)$\times 10^{7}$\\
                &	             &        &	4d5s&	a $^1$D$_2$ &3296.2  & 468.4 &	E1	&	0.7(4) & $\leq4.6\times 10^{6}$ \\ 
                &	             &        &	4d$^2$&	a $^3$F$_2$ &8003.1  & 600.8 &	E1	&	$\leq0.06$& $\leq3.3\times 10^{3}$ \\
                &               &         &	4d$^2$&	a $^3$F$_3$ &8328.0  & 612.8 &  E1  & 0.99(19) &	1.7(7)$\times 10^{4}$\\
                &               &         &	4d$^2$&	a $^3$P$_1$ &14018.3 & 940.8 &  E1  &	0.911(11)&	4.04(9)$\times 10^{5}$\\
                &               &         &	4d$^2$&	a $^3$P$_2$ &14098.1 & 948.0 &  E1  &	1.439(19)&	9.85(26)$\times 10^{5}$\\
                &               &         &	4d$^2$&	b $^1$D$_2$ &14832.9  &1018.9&  E1  &  0.490(11) &	9.2(4)$\times 10^{4}$\\
4d$^2$	        &	  $^1$S$_0$	& 25070.3 &	4d5s&	a $^3$D$_1$ & 840.2 &  412.7 &	M1	&	3$\times 10^{-5}$ & 3.45$\times 10^{-7}$ \\
                &	             &        &	4d5s&	a $^3$D$_2$ &1045.1  & 416.2 &	E2	&	0.421(10) & 1.59(8)$\times 10^{-1}$ \\
                &	             &        &	4d5s&	a $^1$D$_2$ &3296.2  & 459.3 &	E2	&	4.50(15) & 11.1(7) \\
                &	             &        &	4d$^2$&	a $^3$F$_2$ &8003.1  & 585.9 &	E2	&	0.4003(3)& 2.598(4)$\times 10^{-2}$ \\
                &               &         &	4d$^2$&	a $^3$P$_1$ &14018.3 & 904.8 &  M1  &	-0.0633(9)&	1.46(4)$\times 10^{-1}$\\
                &               &         &	4d$^2$&	a $^3$P$_2$ &14098.1 & 911.4 &  E2  &	-4.19(12)&	3.13(18)$\times 10^{-1}$\\
                &               &         &	4d$^2$&	b $^1$D$_2$ &14832.9 & 976.8 &  E2  &  11.23(13) &	1.59(4) \\
                &               &         &	5s5p  &z $^3$P$^o_1$&23776.2 & 7727.9&  E1  &  0.121(11) &	64(12) \\
4d5p            &	z $^1$D$^o_2$&26147.3 &	4d5s&	a $^3$D$_1$ & 840.2 &  395.1 &	E1	&	1.75(7)& 2.02(17)$\times 10^{7}$ \\
                &	             &        &	4d5s&	a $^3$D$_2$ &1045.1  & 398.4 &	E1	&	1.91(9) & 2.34(22)$\times 10^{7}$ \\
                &                &        &	4d5s&	a $^3$D$_3$ &1449.8  & 404.9 &  E1  &   0.18(10)& $\leq4.1\times 10^{5}$\\
                &	             &        &	4d5s&	a $^1$D$_2$ &3296.2  & 437.6 &	E1	&	4.59(11) & 1.02(5)$\times 10^{8}$ \\
                &	             &        &	4d$^2$&	a $^3$F$_2$ &8003.1  & 551.1 &	E1	&	1.13(4)& 3.07(24)$\times 10^{6}$ \\
                &               &         &	4d$^2$&	a $^3$F$_3$ &8328.0  & 561.2 &  E1  & 0.11(4) &	2.8(2.1)$\times 10^{4}$\\
                &               &         &	4d$^2$&	a $^3$P$_1$ &14018.3 & 824.5 &  E1  &	0.09(8)&	$\leq1.7\times 10^{4}$\\
                &               &         &	4d$^2$&	a $^3$P$_2$ &14098.1 & 829.9 &  E1  &	0.16(12)&	$\leq4.6\times 10^{4}$\\
                &               &         &	4d$^2$&	b $^1$D$_2$ &14832.9  &883.8&  E1  &  0.497(8) &	1.45(5)$\times 10^{5}$\\
4d5p            &	z $^3$F$^o_2$&27227.0 &	4d5s&	a $^3$D$_1$ & 840.2 &  379.0 &	E1	&	3.35(4)& 8.35(21)$\times 10^{7}$ \\
                &	             &        &	4d5s&	a $^3$D$_2$ &1045.1  & 381.9 &	E1	&	1.215(23) & 1.07(40)$\times 10^{7}$ \\
                &                &        &	4d5s&	a $^3$D$_3$ &1449.8  & 387.9 &  E1  &   0.616(15)& 2.63(13)$\times 10^{6}$\\ 
                &	             &        &	4d5s&	a $^1$D$_2$ &3296.2  & 417.9 &	E1	&	2.85(6) & 4.53(20)$\times 10^{7}$ \\
                &	             &        &	4d$^2$&	a $^3$F$_2$ &8003.1  & 520.2 &	E1	&	2.170(29)& 1.355(36)$\times 10^{7}$ \\
                &               &         &	4d$^2$&	a $^3$F$_3$ &8328.0  & 529.1 &  E1  & 0.513(8) &	7.21(21)$\times 10^{5}$\\
                &               &         &	4d$^2$&	a $^3$P$_1$ &14018.3 & 757.1 &  E1  &	0.094(11)&	8.2(2.0)$\times 10^{3}$\\
                &               &         &	4d$^2$&	a $^3$P$_2$ &14098.1 & 761.7 &  E1  &	0.138(26)&	1.7(7)$\times 10^{4}$\\
                &               &         &	4d$^2$&	b $^1$D$_2$ &14832.9  &806.8 &  E1  &  0.364(26) &	1.02(14)$\times 10^{5}$\\
4d5p            &	z $^1$P$^o_1$&27516.7 &	5s$^2$&	a $^1$S$_0$	&	0.0 & 363.4	 &	E1	&	2.7(3) & 1.02(23)$\times 10^{8}$ \\
                &	             &        &	4d5s&	a $^3$D$_1$ & 840.2 &  374.9 &	E1	&	1.2(4)& 1.7(1.1)$\times 10^{7}$ \\
                &	             &        &	4d5s&	a $^3$D$_2$ &1045.1  & 377.8 &	E1	&	1.28(22) & 2.0(7)$\times 10^{7}$ \\
                &                &        &	4d5s&	a $^1$D$_2$ &3296.2  & 412.9 &  E1  &   0.07(5)& $\leq1.3\times 10^{5}$\\
                &	             &        &	4d$^2$&	a $^3$F$_2$ &8003.1  & 512.5 &	E1	&	1.5(4)& 1.2(6.0)$\times 10^{7}$ \\
                &	            &         &	4d$^2$&	a $^3$P$_0$	& 13883.4& 733.5&  E1  &   0.34(8)& 2.0(1.0)$\times 10^{5}$ \\
                &               &         &	4d$^2$&	a $^3$P$_1$ &14018.3 & 740.8 &  E1  &	0.29(9)&	1.4(8)$\times 10^{5}$\\
                &               &         &	4d$^2$&	a $^3$P$_2$ &14098.1 & 745.2 &  E1  &	0.77(9)&	9.7(2.2)$\times 10^{5}$\\
                &               &         &	4d$^2$&	b $^1$D$_2$ &14832.9  &788.4 &  E1  &  2.20(22) &	6.7(1.3)$\times 10^{6}$\\
                &               &         &	4d$^2$&	$^1$S$_0$	& 25070.3 &4087.6&  E1  &  0.83(8) &	6.8(1.3)$\times 10^{3}$\\
4d5p            &	z $^3$F$^o_3$&27532.3 &	4d5s&	a $^3$D$_2$ &1045.1  & 377.5 &	E1	&	4.50(6)& 1.09(3)$\times 10^{8}$ \\
                &                &        &	4d5s&	a $^3$D$_3$ &1449.8  & 383.4 &  E1  &   2.30(5)& 2.73(11)$\times 10^{7}$\\
                &                &        &	4d5s&	a $^1$D$_2$ &3296.2  & 412.6 &  E1  &   0.66(4)& 1.78(19)$\times 10^{6}$\\
                &	             &        &	4d$^2$&	a $^3$F$_2$ &8003.1  & 512.1 &	E1	&	0.828(18)& 1.48(6)$\times 10^{6}$ \\
                &               &         &	4d$^2$&	a $^3$F$_3$ &8328.0  & 520.7 &  E1  & 2.83(4) &	1.65(5)$\times 10^{7}$\\
                &               &         &	4d$^2$&	a $^3$F$_4$ &8743.3  & 532.2 &  E1  &0.27(4) &	1.4(5)$\times 10^{5}$\\
                &               &         &	4d$^2$&	a $^3$P$_2$ &14098.1 & 744.4 &  E1  &	0.205(17)&	2.9(5)$\times 10^{4}$\\
                &               &         &	4d$^2$&	b $^1$D$_2$ &14832.9  &787.4 &  E1  &  0.040(15) &	9(7)$\times 10^{2}$\\
                &               &         &	4d$^2$&	a $^1$G$_4$ &15682.9  &843.9 &  E1  &  0.065(8) &	2.0(5)$\times 10^{3}$\\

4d5p            &	z $^3$F$^o_4$&28394.2 &	4d5s&	a $^3$D$_3$ &1449.8  & 371.1 &  E1  &   5.77(7)& 1.47(3)$\times 10^{8}$\\
                &               &         &	4d$^2$&	a $^3$F$_3$ &8328.0  & 498.4 &  E1  & 0.815(14) &	1.21(4)$\times 10^{6}$\\
                &               &         &	4d$^2$&	a $^3$F$_4$ &8743.3  & 508.9 &  E1  &3.21(5) &	1.76(6)$\times 10^{7}$\\
                &               &         &	4d$^2$&	a $^1$G$_4$ &15682.9  &786.7 &  E1  &  0.1139(16) &	5.99(17)$\times 10^{3}$\\
4d5p            &	z $^3$D$^o_1$&28595.3 &	5s$^2$&	$^1$S$_0$	&	0.0 & 349.7 &	E1	&	1.5(4) & 3.5(1.8)$\times 10^{7}$ \\
                &	             &        &	4d5s&	a $^3$D$_1$ & 840.2 &  360.3 &	E1	&	2.68(21)& 1.04(16)$\times 10^{8}$ \\
                &	             &        &	4d5s&	a $^3$D$_2$ &1045.1  & 363.0 &	E1	&	1.43(23) & 2.9(9)$\times 10^{7}$ \\
                &                &        &	4d5s&	a $^1$D$_2$ &3296.2  & 395.3 &  E1  &   0.35(3)& 1.36(22)$\times 10^{6}$\\                
                &	             &        &	4d$^2$&	a $^3$F$_2$ &8003.1  & 485.6 &	E1	&	2.62(29)& 4.0(9)$\times 10^{7}$ \\
                &	            &         &	4d$^2$&	a $^3$P$_0$	& 13883.4& 679.7&  E1  &   0.63(6)& 8.5(1.6)$\times 10^{5}$ \\
                &               &         &	4d$^2$&	a $^3$P$_1$ &14018.3 & 686.0 &  E1  &	0.51(5)&	5.5(1.0)$\times 10^{5}$\\
                &               &         &	4d$^2$&	a $^3$P$_2$ &14098.1 & 689.8 &  E1  &	0.60(14)&	7(3)$\times 10^{5}$\\
                &               &         &	4d$^2$&	b $^1$D$_2$ &14832.9  &726.6 &  E1  &  1.2(3) &	2.7(1.4)$\times 10^{6}$\\
                &               &         &	4d$^2$&	$^1$S$_0$	& 25070.3 &2836.9&  E1  &  0.48(13) &	7(4)$\times 10^{3}$\\
4d5p            &	z $^3$D$^o_2$&28730.0 &	4d5s&	a $^3$D$_1$ & 840.2 &  358.6 &	E1	&	2.031(26)& 3.63(9)$\times 10^{7}$ \\
                &	             &        &	4d5s&	a $^3$D$_2$ &1045.1  & 361.2 &	E1	&	3.37(5) & 9.8(3)$\times 10^{7}$ \\
                &                &        &	4d5s&	a $^3$D$_3$ &1449.8  & 366.6 &  E1  &   2.033(26)& 3.4(9)$\times 10^{7}$\\ 
                &                &        &	4d5s&	a $^1$D$_2$ &3296.2  & 393.2 &  E1  &   0.57(5)& 2.2(4)$\times 10^{6}$\\
                &	             &        &	4d$^2$&	a $^3$F$_2$ &8003.1  & 482.5 &	E1	&	1.095(21)& 4.33(16)$\times 10^{6}$ \\
                &               &         &	4d$^2$&	a $^3$F$_3$ &8328.0  & 490.1 &  E1  & 3.74(6) &	4.82(16)$\times 10^{7}$\\
                &               &         &	4d$^2$&	a $^3$P$_1$ &14018.3 & 679.7 &  E1  &	1.094(17)&	1.55(5)$\times 10^{6}$\\
                &               &         &	4d$^2$&	a $^3$P$_2$ &14098.1 & 683.4 &  E1  &	0.455(18)&	2.63(21)$\times 10^{5}$\\
                &               &         &	4d$^2$&	b $^1$D$_2$ &14832.9  &719.6 &  E1  &  0.215(17) &	5.0(8)$\times 10^{4}$\\
4d5p            &	z $^3$D$^o_3$&29214.0 &		4d5s&	a $^3$D$_2$ &1045.1  & 355.0 &	E1	&	2.37(5) & 3.62(15)$\times 10^{7}$ \\
                &                &        &	4d5s&	a $^3$D$_3$ &1449.8  & 360.2 &  E1  &   4.71(7)& 1.37(4)$\times 10^{8}$\\
                &                &        &	4d5s&	a $^1$D$_2$ &3296.2  & 385.8 &  E1  &   $\leq0.028$& $\leq1.5\times 10^{3}$\\ 
                &	             &        &	4d$^2$&	a $^3$F$_2$ &8003.1  & 471.5 &	E1	&	0.097(7)& 2.6(4)$\times 10^{4}$ \\
                &               &         &	4d$^2$&	a $^3$F$_3$ &8328.0  & 478.8 &  E1  & 0.95(4) &	2.37(19)$\times 10^{6}$\\
                &               &         &	4d$^2$&	a $^3$F$_4$ &8743.3  & 488.5 &  E1  &4.52(8) &	5.07(17)$\times 10^{7}$\\
                &               &         &	4d$^2$&	a $^3$P$_2$ &14098.1 & 661.6 &  E1  &	1.380(20)&	1.9(5)$\times 10^{6}$\\
                &               &         &	4d$^2$&	b $^1$D$_2$ &14832.9  &695.4 &  E1  &  0.357(12) &	1.09(8)$\times 10^{5}$\\
                &               &         &	4d$^2$&	a $^1$G$_4$ &15682.9  &739.0 &  E1  &  0.333(5) &	7.97(23)$\times 10^{4}$\\

\end{longtable*}
\endgroup

\twocolumngrid

\section{Quantum operations}
\label{appendix:quantum_ops}
\subsection{Master equation}

To estimate the impacts of spontaneous emission (photon scattering) during the
Raman shelving and gate operations, we extend the treatments of Refs.~\cite{Uys2010,Moore2023}. Ref.~\cite{Moore2023} evaluated photon scattering rates using Fermi's Golden Rule
to obtain Raman transition rates due to laser photon absorption and emission
into the vacuum, including the effects of counter-rotating terms. Ref. ~\cite{Uys2010}
evaluated dephasing of a two-level system (a qubit) due to Rayleigh
scattering, without considering counter-rotating scattering processes.

To capture dephasing effects between all four states in the nuclear spin
shelving process, and to include the effects of counter-rotating terms in the
dephasing, we derive the two-photon dissipator for the master equation in a
general framework including an arbitrary number of lasers and internal levels,
and accounting for the rotational structure of the problem. We then perform
time-dependent master equation simulations (with the internal degrees of
freedom only) to evaluate the effects of Raman and Rayleigh scattering
processes. Finally, we evaluate the average infidelity of the process, and
extract terms that correspond to population transfer and dephasing errors.

\subsubsection{Interactions}

The interaction of an ion with the vacuum electromagnetic field modes can be
written as
\[ \hat{V}_R (t) = \sum_{\tmmathbf{k}, \alpha} \sum_{k, i = 1}^{N_s}
   V^{(\tmmathbf{k} \alpha)}_{k i} (\hat{\tmmathbf{R}}) \sigma^+_{k i}
   a_{\tmmathbf{k} \alpha} e^{- i (\omega_{\tmmathbf{k}} - \omega_{k i}) t} +
   h.c. \]
where $a_{\tmmathbf{k} \alpha}$ annihilates a photon of momentum $\hbar
\tmmathbf{k}$ and with polarization state $\alpha = 1, 2$; $\sigma_{k i}^+ = |
k \nobracket \rangle \langle \nobracket i |$ is a transition operator between
states $k$ and $i$; and $V_{k i}^{(\tmmathbf{k} \alpha)}$ is a transition
matrix element. In general the state $\ket{i}$ can include motion. Considering
only dipole transitions, one has
\[ V_{k i}^{(\tmmathbf{k} \alpha)} (\hat{\tmmathbf{R}}) = \langle k |
   \hat{\tmmathbf{d}} \cdot \tmmathbf{\varepsilon}_{\tmmathbf{k} \alpha}
   e^{i\tmmathbf{k} \cdot \hat{\tmmathbf{R}}} | i \rangle \nobracket \]
where $\hat{\tmmathbf{d}}$ is the electric dipole operator for the ion and
$\tmmathbf{\varepsilon}_{\tmmathbf{k} \alpha}$ the polarization vector of the
outgoing scattered photon.

The interaction of the ion with the laser fields can be written as
\[ \hat{V}_L (t) = \sum_{\ell} \sum_{k, j = 1}^{N_s} \hat{V}_{k
   j}^{(\ell)} (\hat{\tmmathbf{R}}, t) + h.c. \]
where $\ell$ indexes the laser fields, and (again for dipole
transitions)
\begin{equation}
\begin{array}{lll}
\hat{V}_{k j}^{(\ell)} (\hat{\tmmathbf{R}}, t) & = &V_{k
   j}^{(\ell)}(\hat{\tmmathbf{R}}) e^{-i (\omega^{(\ell)} - \omega_{k j}) t} \sigma_{k j}^+,\\[2mm]
   
   V_{k j}^{(\ell)}(\hat{\tmmathbf{R}}) & = & \langle k | \hat{\tmmathbf{d}}
   \cdot \tmmathbf{\varepsilon}^{(\ell)} e^{i\tmmathbf{k} \cdot
   \hat{\tmmathbf{R}}} | j \rangle 
   \end{array}
   \end{equation}   
where $\tmmathbf{\varepsilon}^{(\ell)}$ is the (possibly complex) polarization vector of laser
$\ell$, $\omega^{(\ell)}$ its angular frequency, and $\omega_{k
j} = \omega_k - \omega_j$ is the angular transition frequency between states
$k$ and $j$.

\subsubsection{Single-photon master equation}

We first consider the master equation for a multilevel ion interacting with
the electromagnetic vacuum via single-photon decays. This single-photon master equation is written as
\begin{equation}
    \frac{d\rho}{dt} = \frac{1}{i} [\hat{H}(t),\rho] + \mathcal{D} \rho
\end{equation}
where $\hat H(t)$ is the Hamiltonian operator including single-photon interactions between the laser fields and the ion, along with terms describing the trapping potential and kinetic energy of the atom.
Typically, the dissipative term (for internal motion only) is written as
\begin{equation}
  \mathcal{D} \rho = \sum_{k > i} \frac{\Gamma_{k i}}{2}  (2 \sigma_{k i}^-
  \rho \sigma^+_{k i} - \sigma^+_{k i} \sigma^-_{k i} \rho - \rho \sigma^+_{k
  i} \sigma_{k i}^-) \label{eq:Dsimple}
\end{equation}
where $i, k$ run over internal states $1$ to $N_s$ in order of increasing
energy, and $\Gamma_{k i}$ is the decay rate from state $k$ to $i$. (The
restriction $k > i$ indicates that only decays from higher- to lower-energy
states are permitted.) However, for a multilevel atom with rotational structure, this form neglects
off-diagonal terms which can potentially contribute to dephasing, and one can
show it breaks rotational invariance in the limit when the bias magnetic field
goes to zero.

A derivation following the standard methods (e.g., Ref.~\cite{Carmichael1}), but
accounting for the full rotational structure of an atom, yields the more
general expression
\begin{equation}
  \begin{array}{lll}
    \mathcal{D} \rho & = & \displaystyle \frac{1}{2} \sum_{i, j} \sum_{k > l} \{ (\gamma_{i
    j, k l} (\omega_{k l}) + \gamma_{i j, k l} (\omega_{i j})) \sigma^-_{k l}
    \rho \sigma^+_{i j} \nobracket\\[4pt]
    &  & \displaystyle \hspace{2em} \left. - \gamma_{i j, k l} (\omega_{k l}) \sigma^+_{i
    j} \sigma^-_{k l} \rho - \gamma_{i j, k l} (\omega_{i j}) \rho \sigma^+_{i
    j} \sigma^-_{k l} \; \right\}
  \end{array}
\end{equation}
where $\gamma_{i j, k l} (\nu)$ are frequency-dependent decay coefficients given below.

In the interaction picture with respect to the state energies, the operators
$\sigma^-_{k l}$, $\sigma^+_{i j}$ acquire a combined time dependence $e^{i
(\omega_{i j} - \omega_{k l}) t}$. Applying a rotating wave approximation, one can 
neglect terms in which $\omega_{i j} - \omega_{k l}$ is large, allowing one
to set $\omega^3_{i j} = \omega^3_{k l}$ in the decay coefficients (for optical transitions). Generally, relatively slow time dependence should be kept and
accounted for by keeping terms for which $| \omega_{i j} - \omega_{k l} | <
\delta$, for some small cutoff $\delta$ (perhaps on the order of MHz), at the
time of performing calculations. Such slow time dependence is consistent with
the approximation $\omega^3_{i j} \approx \omega^3_{k l}$, assuming optical
transition frequencies. The presence of such time-dependent terms with small enough difference frequencies to be non-negligible will depend
on the detailed multilevel structure of the atom and the strength of the bias
field applied. We retain these terms in the formalism to allow any accidental degeneracies to be captured in our calculations.

One thus obtains
\begin{equation}
  \mathcal D \rho = \!\!\sum_{i > j, \, k > l} \!\!\!
  \frac{\gamma_{i j, k l} (\omega_{k l})}{2}  (2 \sigma^-_{k l} \rho
  \sigma^+_{i j} - \sigma^+_{i j} \sigma^-_{k l} \rho - \rho \sigma^+_{i
  j} \sigma^-_{k l})  \label{eq:2ndOrderMaster}
\end{equation}
where the frequency-dependent decay coefficients $\gamma_{i j, k l} (\nu)$ are
given by
\begin{equation}
\gamma_{i j, k l} (\nu) = \frac{2 \pi}{\hbar^2} \sum_{\tmmathbf{k}, \alpha}
   V_{i j}^{(\tmmathbf{k} \alpha)} (\hat{\tmmathbf{R}}) V^{(\tmmathbf{k}
   \alpha) \ast}_{k l} (\hat{\tmmathbf{R}}) \delta (\omega_{\tmmathbf{k}} -
   \nu)
\end{equation}
and where the sum is over vacuum modes and polarizations contained in some
volume. Taking this volume to be arbitrarily large, approximating the sum as
an integral, and evaluating for dipole transitions, yields
\begin{widetext}
\begin{equation}
    \gamma_{i j, k l} (\nu) = \frac{\nu^3}{3 \pi \hbar c^3 \varepsilon_0} 
    \elem{i}{\hat{d}^{(1)}_{m_{i j}}}{j}  \elem{k}{\hat{d}^{(1)}_{m_{k
    l}}}{l}^{\ast}
     \;\frac{3}{8 \pi} \sum_{\alpha} \int d^2 \tmmathbf{u} \;
    \epsilon^{(1)}_{\tmmathbf{u} \alpha, m_{i j}}  \langle \tmmathbf{n}_i
    |e^{i k\tmmathbf{u} \cdot \hat{\tmmathbf{R}}} |\tmmathbf{n}_j \rangle
    \epsilon^{(1) \ast}_{\tmmathbf{u} \alpha, m_{k l}}  \langle \tmmathbf{n}_k
    |e^{i k\tmmathbf{u} \cdot \hat{\tmmathbf{R}}} |\tmmathbf{n}_l
    \rangle^{\ast}
  \end{equation}
  \end{widetext}
where $\nu = k c$, $\elem{i}{\hat{d}^{(1)}_{m_{i j}}}{j}$ is the internal dipole transition matrix
element between states $i$ and $j$, $m_{i j} = m_i - m_j$ is the
difference in magnetic quantum numbers, and $\hat{d}^{(1)}_m$ are the
spherical components of the dipole operator (the superscript ${(1)}$ is used to indicate that the subscript indexes the $L=1$ spherical components instead of Cartesian coordinates). The integral is over all unit direction vectors $\tmmathbf{u}$ of the emitted photon. Here $\tmmathbf{n}_i$ is the motional state associated with state $i$, which for trapped ions is
typically a list of Fock state occupations $n_{i, \mu}$ for each normal mode
$\mu = 1, \ldots, N_{\tmop{modes}}$ of low-energy harmonic motion in an ion crystal. (Note we have abused notation slightly in re-using $i$ to denote both the index of an internal state and a state including the internal motion, $\ket{i} = \ket{i,\tmmathbf{n}_i}$. When a matrix element, e.g., $\elem{i}{\hat{d}^{(1)}_{m_{i j}}}{j}$, involves only operators acting on the internal degrees of freedom, it can be understood that the indices $i$, $j$ refer to the internal states.)

When the external motion is not relevant, one can set $\hat{\tmmathbf{R}} = 0$ to obtain
\begin{equation}
    \label{eq:gamNoMotion}
\gamma_{i j, k l} (\nu) = \delta_{m_{i j}, m_{k l}}  \frac{\nu^3}{3 \pi \hbar c^3 \varepsilon_0}  
   \elem{i}{\hat{d}^{(1)}_{m_{i j}}}{j}   \elem{k}{\hat{d}^{(1)}_{m_{k l}}}{l} .
\end{equation}
The coefficients $\gamma_{i j, k l} (\nu)$ enter into the two-photon
dissipator below. Note that the Kronecker $\delta$ applies only to the
magnetic quantum number differences $m_{i j}$, $m_{k l}$. This yields a
different structure than Eq. \eqref{eq:Dsimple}, in which $\delta_{m_{i j}, m_{k l}}$ in Eq.~\eqref{eq:gamNoMotion} is replaced with with $\delta_{i, k} \delta_{j, l}$, thereby omitting off-diagonal terms.

\subsubsection{Effective coherent Raman interaction}
For Raman transitions, the single-photon detunings are very large compared to any other frequency scales in the dynamics, and it is appropriate to adiabatically eliminate the excited states to obtain an effective two-photon interaction describing the dynamics only in the low-energy space of initial and final states. For this we extend the treatment of Ref.~\cite{James2007} to include counter-rotating terms, which gives, in the interaction picture,
\begin{widetext}
\begin{equation}
  \begin{array}{lll}
    \hat{V}_I (t) & = & \displaystyle \sum_{\ell', \ell} \sum_{i, k, j} \left(
    \frac{V_{k, i}^{(\ell') *} (\hat{\tmmathbf{R}}, t) V_{k,
    j}^{(\ell)} (\hat{\tmmathbf{R}}, t)}{\omega^{(\ell)} - (\omega_k -
    \omega_j)} - \frac{V_{i, k}^{(\ell)} (\hat{\tmmathbf{R}}, t) V_{j,
    k}^{(\ell') *} (\hat{\tmmathbf{R}}, t)}{\omega^{(\ell)} - (\omega_i
    - \omega_k)} \right)  \hat{\sigma}_{i j} (t) + h.c.
  \end{array}
\end{equation}
Here $k$ ranges over the intermediate (excited) states in the Raman transitions, while $i,j$ ranges over the low-energy states.
\subsubsection{Effective Raman scattering interaction}

One can use time-dependent perturbation theory to derive an effective
interaction for the two-photon process of absorption and spontaneous emission,
similar to Ref.~\cite{Uys2010}, but including counter-rotating terms. We omit details of the derivation, but the result is, in
the interaction picture,
\begin{equation}
  \hat{V}_I(t) = \sum_{\ell} \sum_{\tmmathbf{k}, \alpha} \sum_{i, j, k}
  \left\{ \frac{V^{(\tmmathbf{k} \alpha) \ast}_{k i} (\hat{\tmmathbf{R}}) V_{k
  j}^{(\ell)} (\hat{\tmmathbf{R}}, t)}{\omega^{(\ell)} -
  \omega_{k j}}  \left. - \frac{V_{i k}^{(\ell)}
  (\hat{\tmmathbf{R}}, t) V^{(\tmmathbf{k} \alpha) \ast}_{j k}
  (\hat{\tmmathbf{R}})}{\omega_{\tmmathbf{k}} + \omega_{k j}}  \right\}
  a^{\dag}_{\tmmathbf{k} \alpha} (t) \sigma^+_{i j} (t) + h.c. \right.
  \label{eq:V2eff}
\end{equation}
\end{widetext}
where $a^{\dag}_{\tmmathbf{k} \alpha} (t) = a^{\dag}_{\tmmathbf{k} \alpha}
e^{i \omega_{\tmmathbf{k}} t}$ and $\sigma^+_{i j} (t) = \sigma^+_{i j} e^{i
\omega_{i j} t}$. \ This interaction describes Raman scattering between two
states $i, j$ through an intermediate state $k$ in which one photon is
absorbed from the laser and one is emitted into the vacuum. The first term
describes absorption of a photon from the laser at frequency
$\omega^{(\ell)}$ while transitioning from $j \rightarrow k$, followed
by emission of a photon of frequency $\omega_{\tmmathbf{k}}$ into the vacuum and
transitioning from $k \rightarrow i$. The second (counter-rotating) term
describes emission a photon into the vacuum while transitioning $j
\rightarrow k$, followed by absorption of a photon from the laser field and
transitioning $k \rightarrow i$. By conservation of energy, one can take
\[ \omega_{\tmmathbf{k}} + \omega_{k j} = \omega^{(\ell)} + \omega_{k
   i} \]
in the demoninator above.

\subsubsection{Master equation for internal motion}

The interaction \eqref{eq:V2eff} can be used to derive a dissipator describing
the effects of spontaneous emission, similar to the approach of Ref.~\cite{Uys2010},
using the 2nd-order Dyson series in the interaction picture,
\begin{eqnarray}
  \frac{\partial \rho_I (t)}{d t} & = & - \int_{t_0}^t d t' \tmop{tr}_R 
  [\hat{V}_I (t), [\hat{V}_I (t'), \rho (t') R]] \label{eq:Dyson2} 
\end{eqnarray}
Inserting $\hat{V}_I$, one applies the usual Born and Markov approximations,
traces over the vacuum modes, integrates out the photon correlation functions
involving pairs of operators $a^{\dag}_{\tmmathbf{k} \alpha} (t)
a_{\tmmathbf{k} \alpha} (t)$ or $a _{\tmmathbf{k} \alpha} (t)
a^{\dagger}_{\tmmathbf{k} \alpha} (t)$, assumes the vacuum photon
distributions are zero-temperature, and finally applies a rotating wave
approximation. The result is
\begin{widetext}
\begin{equation}
  \begin{array}{lll}
    \displaystyle \frac{\partial \rho_I (t)}{d t} \; = \; \mathcal{D}_{\mathrm{eff}}(\rho_I,t) & = &  \displaystyle \frac{1}{2} \sum_{i' j', i j}
    \sum_{\ell', \ell} e^{i ([\omega^{(\ell')} -
    \omega_{i' j'}] - [\omega^{(\ell)} - \omega_{i j}]) t}
    \gamma^{(\ell', \ell)}_{i' j', i j} (\omega^{(\ell')}
    - \omega_{i' j'})\\[4pt]
    &  & \displaystyle \qquad \qquad \qquad \times [2 \sigma_{i j}^+ \rho_I (t) \sigma_{i'
    j'}^- - \rho_I (t) \sigma_{i' j'}^- \sigma_{i j}^+ - \sigma_{i' j'}^-
    \sigma_{i j}^+ \rho_I (t)]\\
  \end{array}  \label{eq:EffectiveDissipator}
\end{equation}
Here the rotating-wave approximation has been applied to keep only terms for which
$\omega^{(\ell')} - \omega_{i' j'} \approx \omega^{(\ell)} -
\omega_{i j}$, similar to the rotating-wave approximation applied for the
single-photon dissipator above. As before, time-dependent terms should be kept
up to some appropriate frequency cutoff $\delta$. \ The two-photon decay
coefficients $\gamma^{(\ell', \ell)}_{i' j', i j} (\nu)$ are
given by
    \begin{equation}
  \begin{array}{lll}
    \displaystyle \gamma^{(\ell', \ell)}_{i' j', i j} (\nu) & = & \displaystyle \sum_{k, k'}
    \{ \gamma_{k' i', k i} (\nu) \mathcal{V}^{(\ell') \ast}_{k' j'}
    \mathcal{V}^{(\ell)}_{k j} - \gamma_{k' i', j k} (\nu)
    \mathcal{V}^{(\ell') \ast}_{k' j'} \mathcal{V}^{(\ell)}_{i
    k} \nobracket\\
    &  & \displaystyle \hspace{3em} - \gamma_{j' k', k i} (\nu) \mathcal{V}^{(\ell')
    \ast}_{i' k'} \mathcal{V}^{(\ell)}_{k j} \nobracket + \gamma_{j'
    k', j k} (\nu) \mathcal{V}^{(\ell') \ast}_{i' k'}
    \mathcal{V}^{(\ell)}_{i k} \}\\
  \end{array}  \label{eq:gamlpl}
\end{equation}
\end{widetext}
where the coefficients $\mathcal{V}^{(\ell)}_{k j}$ are given by
\begin{equation}\label{eq:scriptV}
  \mathcal{V}^{(\ell)}_{k j} = V_{k j}^{(\ell)}(\hat{\tmmathbf{R}}) /
  (\omega^{(\ell)} - \omega_{k j}) .
\end{equation}
and where $\gamma_{i j,k l}(\nu)$ are given by Eq.~\eqref{eq:gamNoMotion} . In this work we only include the effect of scattering on the internal degrees of freedom, and therefore set $\hat{\tmmathbf{R}} = \tmmathbf{0}$ in~\eqref{eq:scriptV}. All state indices in Eqs.~\eqref{eq:EffectiveDissipator} and \eqref{eq:gamlpl} are therefore taken to be internal state indices only.

\subsubsection{Scattering rate}

One can check that Eq. \eqref{eq:EffectiveDissipator} recovers the usual scattering
expression derived from Fermi's Golden Rule. Beginning with an atom in state
$\rho^{(m)} = \ket{m} \bra{m}$, the scattering rate $m \rightarrow n \neq m$
is
\begin{equation}
  \Gamma_{m \rightarrow n} = \frac{d \rho^{(m)}_{n, n}}{d t} =
  \elem{n}{\mathcal{D_\mathrm{eff}} \left( \ket{m} \bra{m} \right)}{n} \qquad (m \neq n) .
  \label{eq:drhonn}
\end{equation}
Additionally, $d \rho^{(m)}_{m, m} / d t$ will give the total scattering rate out of $m$. To
conserve probability, one should have
\[ \sum_n \frac{d \rho^{(m)}_{n, n}}{d t} = 0 \]
so that $d \rho^{(m)}_{m, m} / d t = - \sum_{n \neq m} d \rho^{(m)}_{n, n} / d
t $.

Evaluating $\Gamma_{m \rightarrow n}$ by computing the matrix elements of Eq.
\eqref{eq:EffectiveDissipator}, assuming $\omega^{(\ell')} - \omega_{i' j'} =
\omega^{(\ell)} - \omega_{i j}$ to exclude all time dependent terms,
one finds
\begin{widetext}
\[ \begin{array}{lll}
     \displaystyle \Gamma_{m \rightarrow n} & = & \displaystyle \sum_{\ell} \gamma^{(\ell,
     \ell)}_{n m, n m} (\omega^{(\ell)} - \omega_{n m})\\
     \qquad & = & \displaystyle \sum_{\ell} \frac{(\omega^{(\ell)} - \omega_{n
     m})^3}{3 \pi \hbar c^3 \varepsilon_0}  \frac{(E^{(\ell)})^2}{4} 
     \sum_q \left| \sum_k \left( \frac{ \elem{n}{\hat{d}_q^{(1)}}{k}
     V^{(\ell)}_{k m} }{\omega^{(\ell)} - \omega_{k m}} -
     \frac{V^{(\ell)}_{n k}  \elem{k}{\hat{d}_q^{(1)}}{m}}{\omega^{(\ell)} - \omega_{n k}} \right)
     \right|^2\\
   \end{array} \]
   \end{widetext} 
This reproduces exactly the expression given in the appendix of Ref.~\cite{Moore2023}.

\subsection{Gate fidelities}

The average fidelity of a single-qubit quantum logic operation, allowing for the possibility of leakage to states outside the computational (qubit) subspace, can be determined by a simple adaptation of results in Ref.~\cite{Cabrera2007}. (Effects of leakage have also been analyzed in Ref.~\cite{Chen2025}.) Denoting the trace over the computational subspace as $\mathrm{tr}_{\mathcal{C}}$, the fidelity of a gate operation $\mathcal{M}(\rho)$, where $\mathcal{M}$ is a noisy quantum channel intended to implement the ideal unitary $U$, can be written as
\begin{equation} \label{eq:fid1}
F = \frac{1}{4} \mathrm{tr}_{\mathcal{C}} [(\mathbbm{1} + U \tmmathbf{P} U^\dagger) (\mathcal{M}(\mathbbm{1}) + \mathcal{M}(\tmmathbf{P}))] 
\end{equation}
where the initial density matrix is written as $\rho = (\mathbbm{1} + \tmmathbf{P})/2$, where $\tmmathbf{P} = P^1 \sigma_1 + P^2 \sigma_2 + P^3 \sigma_3$ and the real-valued components $(P^1, P^2, P^3)$ form a unit vector. Averaging the input over the unit sphere gives the average fidelity
\begin{equation} \label{eq:avgFidelity}
\langle F \rangle = \frac{1}{4} \mathrm{tr}_{\mathcal{C}} \left( \mathcal{M}(\mathbbm{1}) + \frac{1}{3} \sum_j U \sigma_j U^\dagger \mathcal{M}(\sigma_j) \right)
\end{equation}

We compute errors due to spontaneous emission by time-evolving the Pauli matrices $I$, $X$, $Y$, $Z$ under the effective dissipator~\eqref{eq:EffectiveDissipator} and computing $1-\langle F\rangle$ using Eq.~\eqref{eq:avgFidelity}. This includes errors due to population transfer between qubit states, leakage to spectator states, as well as dephasing (including Rayleigh dephasing). Dephasing effects can be separated out from population transfer effects by looking at the errors associated with the input matrices $X$, $Y$. However, we generally find that dephasing effects are substantially smaller than population transfer errors, and have a similar dependence on wavelength and bias field magnitude, so we report only the total error $1-\langle F\rangle$.
\end{document}